\documentclass[12pt,preprint]{aastex}

\usepackage{color}

\begin{document}

\title{Modeling the Multi-Wavelength Emission of Shell-Type Supernova Remnant
RX J1713.7-3946}

\author{Qiang Yuan$^1$, Siming Liu$^{2,3}$, Zhonghui Fan$^4$,
Xiaojun Bi$^{1,5}$, and Christopher L. Fryer$^{6,7}$}

\affil{
 $^{1}$Key Laboratory of Particle Astrophysics, Institute of High Energy
Physics, Chinese Academy of Sciences, Beijing 100049, P. R. China\\
 $^{2}$Department of Physics and Astronomy, University of Glasgow,
Glasgow G12 8QQ, UK\\
 $^{3}$Key Laboratory of Dark Matter and Space Astronomy, Purple Mountain 
Observatory, Chinese Academy of Sciences, Nanjing 210008, P. R. China\\
 $^{4}$Department of Physics, Yunnan University, Kunming 650091,
Yunnan, P. R. China\\
 $^{5}$Center for High Energy Physics, Peking University, Beijing 100871,
P.R.China \\
 $^{6}$Los Alamos National Laboratories, Los Alamos, NM 87545\\
 $^{7}$Physics Department, University of Arizona, Tucson AZ 85721
}

\begin{abstract}

Emission mechanisms of the shell-type supernova remnant (SNR) RX J1713.7-3946
are studied with multi-wavelength observational data from radio, X-ray,
GeV $\gamma$-ray to TeV $\gamma$-ray band. A Markov Chain Monte Carlo
method is employed to explore the high-dimensional model parameter
space systematically. Three scenarios for the $\gamma$-ray emission are
investigated: the leptonic, the hadronic and a hybrid one. Thermal
emission from the background plasma is also included to constrain the
gas density, assuming ionization equilibrium, and a 2$\sigma$ upper limit
of about $0.03$ cm$^{-3}$ is obtained as far as thermal energies account
for a significant fraction of the dissipated kinetic energy of the SNR shock.
Although systematic errors dominate the $\chi^2$ of the spectral fit of 
all models, we find that 1) the leptonic model has the best constrained 
model parameters, whose values can be easily accommodated with a typical 
supernova, but gives relatively poor fit to the TeV $\gamma$-ray data; 
2) The hybrid scenario has one more parameter than the leptonic one and 
improves the overall spectral fit significantly; 3) The hadronic one, 
which has three more parameters than the leptonic model, gives the best 
fit to the overall spectrum with relatively not-well-constrained model 
parameters and very hard spectra of accelerated particles. The uncertainties 
of the model parameters decrease significantly if the spectral indices of 
accelerated electrons and protons are the same. The hybrid and hadronic 
models also require an energy input into high-energy protons, which seems 
to be too high compared with typical values of a supernova explosion. Further
investigations are required to reconcile these observations with SNR theories.

\end{abstract}


\section{Introduction}

Supernova remnants (SNRs) are widely thought to be one important kind
of cosmic ray (CR) sources in the Galaxy \citep{2004Natur.432...75A}.
The most direct evidence comes from high-energy $\gamma$-ray emission
from SNRs. Generally there are two types of scenarios for production
of high energy $\gamma$-rays: the leptonic (via inverse Compton scattering
of background photos by relativistic electrons) and hadronic (via decay
of neutral pions produced by elastic collisions of relativistic ions with
ions in the background plasma) origins. Understanding which of these two
scenarios is dominant in specific sources is very important for the search
of CR nuclei sources and the study of CR acceleration \citep{Gabici2008}.

Usually it is difficult to distinguish the leptonic model and hadronic
model just with the high energy $\gamma$-ray data alone. Multi-wavelength
observations of photon emission from SNRs can provide us key information
about the radiation mechanism. Shell-type SNR RX J1713.7-3946 is one of the
most widely studied SNRs with perhaps the best multi-wavelength observations.
The observational data span from radio \citep{2004ApJ...602..271L},
infrared \citep{2003PASP..115..953B, 2009A&A...505..157A},
X-ray \citep{1997PASJ...49L...7K,2003A&A...400..567U,2004A&A...427..199C},
GeV $\gamma$-ray \citep{fermi:rxj1713}, to TeV $\gamma$-ray band
\citep{2000A&A...354L..57M,2002Natur.416..823E,2006A&A...449..223A}.
Recent observations, especially the X-ray emission detected by $Suzaku$
\citep{2008ApJ...685..988T} and TeV $\gamma$-ray emission measured by
$HESS$ \citep{2007A&A...464..235A}, give the energy spectra and images
of this SNR with very high quality, which makes detailed modelings of
the emission mechanism plausible
\citep{2009MNRAS.392..240M,2009MNRAS.392..925F,2010MNRAS.406.1337F,
2010ApJ...708..965Z,2010ApJ...712..287E,2010A&A...517L...4F}. The newly
reported data from $Fermi$ \citep{fermi:rxj1713} also set strict
constraints on the nature of the radiation from this SNR.

Basic results of recent studies of this SNR may be summarized briefly as
the following. The wide range TeV $\gamma$-ray spectrum favors a hadronic
origin of the high energy emission \citep{2006A&A...449..223A,
2009A&A...496....1D,2009MNRAS.392..240M,2009MNRAS.392..925F, 
2010A&A...511A..34B}. This scenario is also in line with the long 
standing view that SNRs are the most important CR accelerators 
\citep{1981ICRC...12..155A}. However, there is a strong correlation 
between the X-ray image and TeV $\gamma$-ray image, favoring a leptonic 
origin of the multi-wavelength emission \citep{2006A&A...449..223A,
2009A&A...505..157A}. \cite{2008NewA...13...73P} also claims that the 
lack of spatial correlation between $\gamma$-rays and the molecular 
cloud in the vicinity of SNR RX J1713.7-3946 argues against the hadronic 
scenario. Furthermore the lack of thermal line emission on the X-ray 
spectrum sets an upper limit on the ambient plasma density of about 
$0.02$ cm$^{-3}$ \citep{2004A&A...427..199C}, which implies a very high 
energy content of accelerated protons from the supernova explosion. 
The hadronic model actually has a proton acceleration efficiency more 
than 4 orders of magnitude higher than the electron acceleration 
efficiency, which corresponds to a rather extreme scenario \citep{butt08}.
It has been shown that the leptonic model can reproduce the multi-wavelength 
data and the model parameters can be easily accommodated by typical SNRs
\citep{2010ApJ...708..965Z, 2010ApJ...712..287E} though the overall fit
to the data is relatively poor for the simplest cases. The spectral fit
can be improved by considering details of electron acceleration near the
high energy cutoff \citep{2008ApJ...683L.163L,2010MNRAS.406.1337F,
2010A&A...517L...4F}. In fact, the cutoff of the TeV spectrum at a few 
tens of TeV favors the leptonic model since production of photons at 
even higher energies through inverse Comptonization of the cosmic 
microwave radiation by TeV electrons is in the Klein-Nishina regime and 
therefore very inefficient \citep{Gabici2008}. The observed decay of 
bright X-ray filaments with a width of $\sim 0.1$ lyr on a timescale 
of $\sim1$ year, on the other hand, can be attributed to fast diffusion 
of high energy electrons in a weak magnetic field away from intermittently 
formed regions of relatively high accelerated electron density 
\citep{2007Natur.449..576U,2008ApJ...683L.163L}.

It is evident that even with both theoretical and observational advances 
on this source recently, the nature of the TeV emission is still 
inconclusive. The multi-wavelength observational data justifies a 
systematic modeling of the emission spectrum. Moreover, the thermal 
emission needs to be taken into account to get more quantitative 
constraints. This paper focuses on these two aspects.
\cite{2010A&A...517L...4F} studied the goodness-of-fit for various
physically motivated leptonic models and found that the diffusive shock
acceleration and the stochastic acceleration give comparably good fits.
In this work we generalize this analysis by including the hadronic component.
We employ the Markov Chain Monte Carlo (MCMC) method to constrain the
model parameters, and investigate the full, correlated parameter
space systematically. The non-thermal spectra of CR electrons and/or
protons are parameterized in the simplest way, i.e.,
a power law with a high-energy cutoff. The thermal bremmstrahlung
radiation and line emission of the background plasma are also taken into
account in the fit. We consider three scenarios for the
$\gamma$-ray emission, the purely leptonic model, the hadronic
model and a hybrid model where the number of model parameters is reduced
by requiring the spectral parameters of CR protons and electrons are
identical except the normalization (see Sec. 2. below). The fitting
results are presented in Section 2 and show consistency with previous
studies. The conclusion is drawn in Section 3, where we also discuss
possible future researches necessary to improve our understanding of
this source.

\section{Fitting results}

In this section, we use the MCMC technique to constrain the model parameters.
The MCMC method is well suitable for high dimensional parameter space
investigation. The Metropolis-Hastings algorithm is used when sampling
the model parameters. The probability density distributions of the model
parameters can also be simply approximated by the number density of the
sample points. A brief introduction to the basic procedure of the MCMC
sampling can be found in \cite{2010A&A...517L...4F}. For more details
about the MCMC method, please refer to \cite{Neal1993,Gamerman1997,
2003itil.book.....M}.

We also discuss implications of model parameters from the best fits to
multi-wavelength data of SNR RX J1713.7-3946 for three scenarios of
the $\gamma$-ray emission. In all these scenarios, the radio to X-ray
emissions are generated through synchrotron of relativistic electrons.
The high energy $\gamma$-rays are produced with different mechanisms.
The basic physical parameters of SNR RX J1713.7-3946 are adopted as:
Age $T_{\rm life}\approx 1600$ yr, Distance $d\approx 1$ kpc, and Radius
$R\approx 10$ pc \citep{1997A&A...318L..59W}, and we assume a uniform
emission sphere with a radius of $R$ in deriving related quantities. 
Although the errors of the Fermi data are large, we still include these 
data in the spectral fits \citep{fermi:rxj1713}. The procedure described 
in this paper can be applied to future observations with improved data 
to evaluate different emission models.

\subsection{Leptonic scenario}

In the leptonic scenario the $\gamma$-ray emission is produced through
inverse Compton (IC) scattering of energetic electrons off the background
radiation field, including the interstellar infrared, optical radiation,
and the cosmic microwave background (CMB). The energy spectrum of
accelerated electrons is prescribed as $F_e(E)\propto E^{-\alpha_e}\exp
\left[-(E/E_c^e)^{\delta_e}\right]$, where $E$, $\alpha_e$, $E_c^e$ are
the electron energy, power law spectral index, high-energy cutoff energy,
respectively, and $\delta_e$ describes the sharpness of this cutoff.
The normalization is given through the total energy of electrons above
1 GeV, $W_e$. The synchrotron radiation also depends on the magnetic
field strength $B$.

The interstellar radiation field (ISRF) other than the CMB may be important
for the calculation of IC $\gamma$-ray spectrum. The
inclusion of ISRF has been proposed to improve the fit to the $HESS$ data
\citep{2006ApJ...648L..29P}. However, given the new
data of X-ray by $Suzaku$ and TeV $\gamma$-ray by $HESS$, it was shown
that only if the intensity of ISRF is artificially boosted by more than
one order of magnitude, the goodness-of-fit can be improved significantly
\citep{2008ApJ...685..988T,2009MNRAS.392..240M}. In this work the ISRF
is adopted as that given by \cite{2006ApJ...648L..29P} at a distance of
$7.5$ kpc from the Galactic center and in the equatorial plane of the
Galactic disk \citep{2006ApJ...640L.155M}. Our results are not sensitive
to details of the ISRF.

Thermal X-ray emission mostly depends on the density of the shocked
interstellar medium (ISM) $n_{\rm ISM}$ and the temperature of background
electrons $T_e$. Depending on the effect of accelerated particles on 
the shock structure, the density is a factor of a few higher than the 
density in the un-shocked upstream region \citep{1999ApJ...526..385B, 
2005ApJ...634..376W}. The electron temperature due to Coulomb collisional 
energy exchange with ions is estimated by
\citet{2000ApJ...543L..61H,2010MNRAS.406.1337F}
\begin{equation}
T_e>2.1\times 10^7 \left(\frac{T_{\rm life}}{1600\,{\rm yr}}\right)^{2/5}
\left(\frac{n_{\rm ISM}}{{\rm cm}^{-3}}\right)^{2/5}
\left(\frac{T_i}{1.3\times 10^8\,{\rm K}}\right)^{2/5},
\end{equation}
where $T_i$ is the temperature of background ions. $T_i$ is estimated to
be higher than $\sim 1.3\times 10^8$ K if the background is heated by the
shock \citep{2010MNRAS.406.1337F}. For $n_{\rm ISM}\approx 0.02$ cm$^{-3}$,
$T_e$ should be higher than $4\times 10^6$ K.
\cite{2009A&A...496....1D} alternatively proposed that the
post-shock region temperature could be reduced significantly in the case
of large Mach number of the shock and effective particle acceleration. If
this is the case, as we will show below, the constraint on the density
and therefore the relativistic proton energy in the hadronic and hybrid 
models will be less strict. However, there are still significant 
uncertainties in the plasma heating downstream \citep{2007ApJ...654L..69G, 
Gabici2008}, and it was also argued that such an extreme condition was 
not easy to meet and the post-shock plasma should be heated more strongly
\citep{2010ApJ...712..287E}. Since there is no direct constraint on $T_e$ 
due to the lack of thermal emission, instead of including $T_e$ in the 
MCMC fit, we take some typical values of $T_e$. For the sake of 
simplicity, we assume that the background electrons are
in ionization equilibrium with the background ions and use the
Raymond-Smith plasma code to calculate the thermal emission for given
$n_{\rm ISM}$, $T_e$ and the metalicity \citep{1977ApJS...35..419R}.
The emission includes the recombination, bremmstrahlung, two photon
process, and line emissions. 
The chemical abundance of the ISM is taken from \cite{1973asqu.book.....A}.
For emission lines lying in the energy range of X-ray data ($0.5-33$ keV),
we convolve the model spectrum with a Gaussian energy spread function, whose
width is adopted as the characteristic energy resolution of ${\it Suzaku}$
\citep{2007PASJ...59S..23K}.

\begin{figure}[!htb]
\centering
\includegraphics[width=0.32\columnwidth]{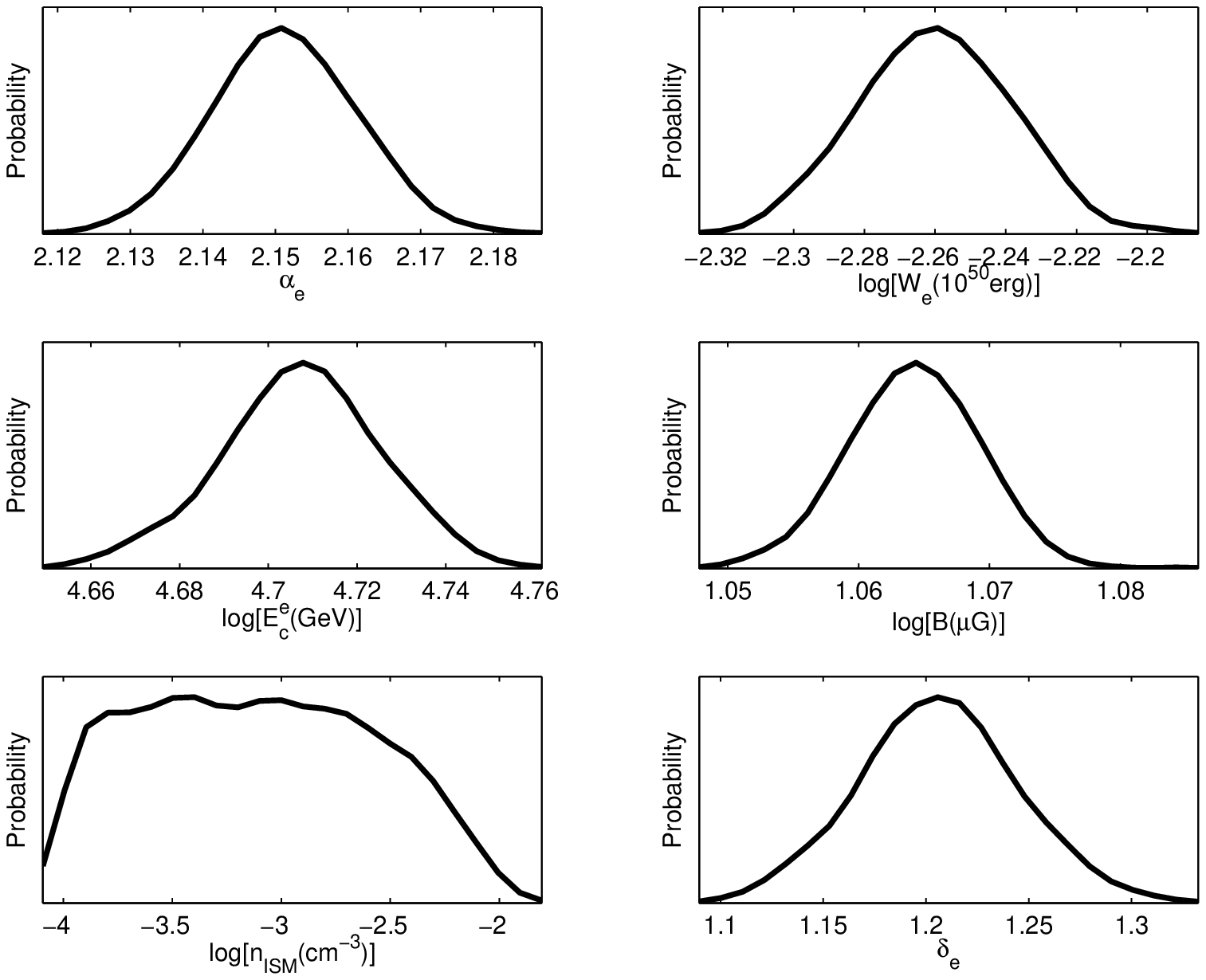}
\includegraphics[width=0.32\columnwidth]{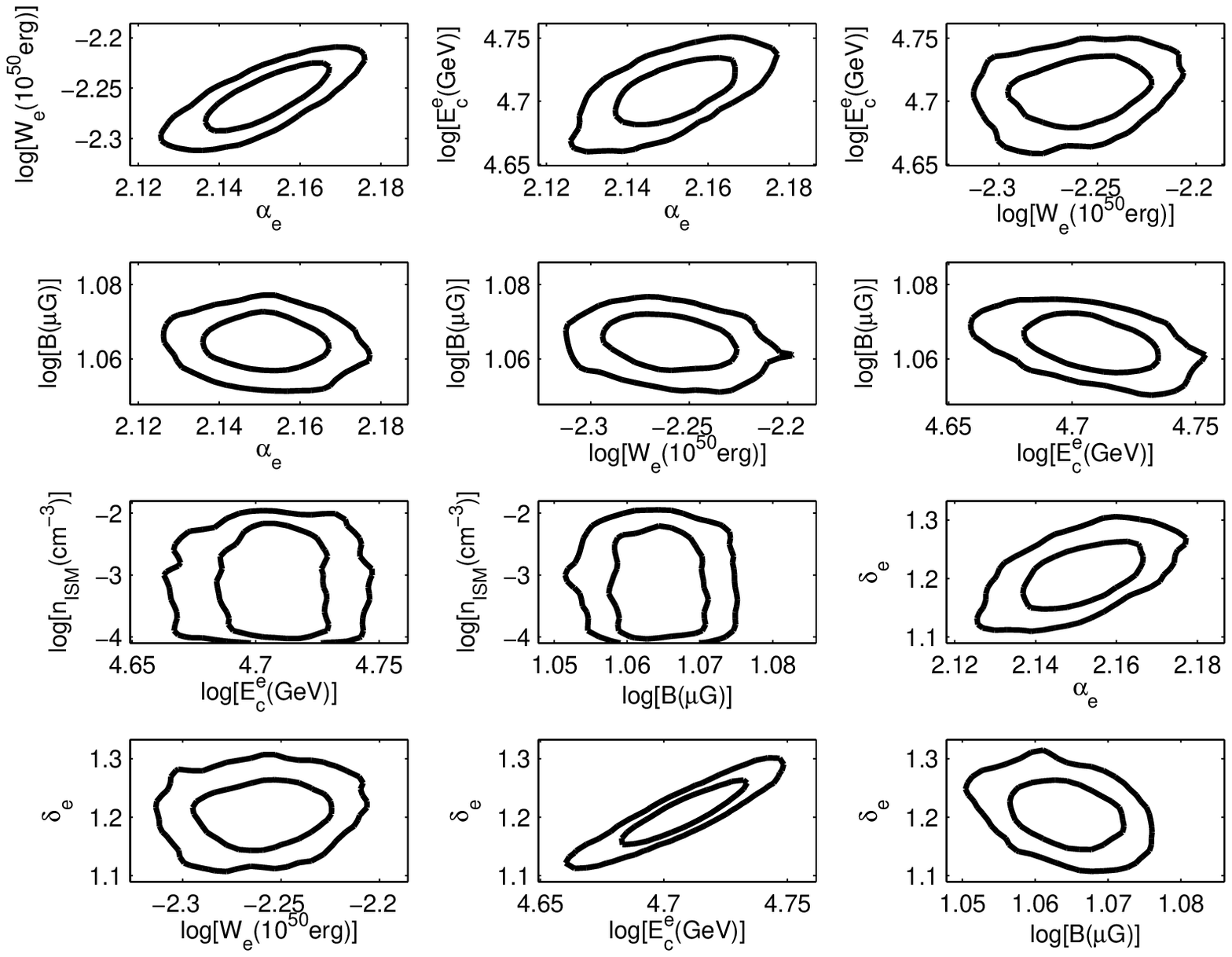}
\includegraphics[width=0.32\columnwidth]{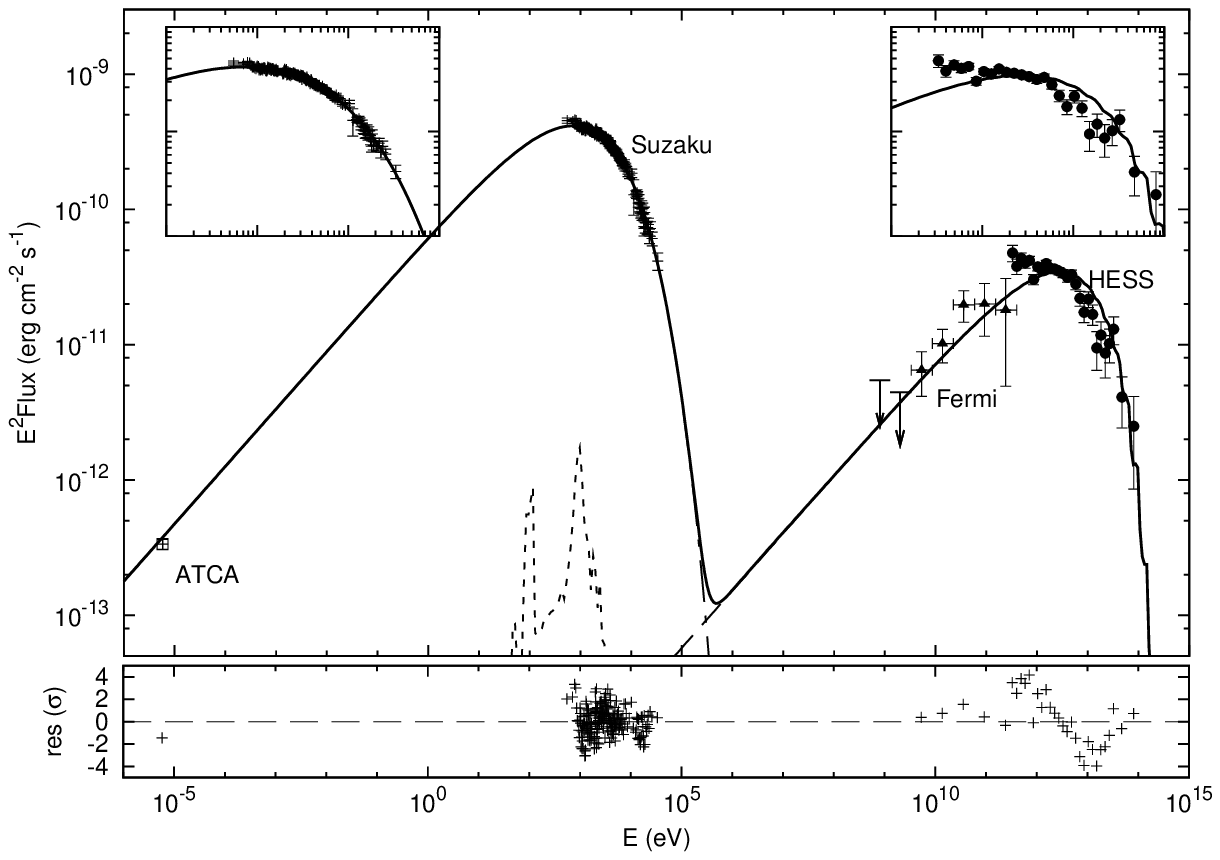}
\caption{Left: 1-D probability distribution of the parameters in the
leptonic model; Middle: 2-D confidence contours of the parameters. The
contours are for 1 and 2 $\sigma$ levels; Right: the best fit to the
spectral energy distribution (SED) from radio \citep{2009A&A...505..157A},
X-ray \citep{2008ApJ...685..988T}, GeV $\gamma$-ray \citep{fermi:rxj1713}
and TeV $\gamma$-ray observations \citep{2007A&A...464..235A}.
The background electron temperature is $10^7$ K. The thermal emission
component indicated by the dotted line corresponds to the $2\sigma$ upper
limit of $0.007$ cm$^{-3}$ for $n_{\rm ISM}$.
}
\label{fig:lepton}
\end{figure}

In total there are 6 free parameters, $\alpha_e,\,
E_c^e,\,W_e,\,\delta_e,\,B,$ and $n_{\rm ISM}$ in the lepton model.
The 1-dimensional (1-D) probability distributions and 2-dimensional (2-D)
confidence regions (at $1\sigma$ and $2\sigma$ confidence levels) of the
model parameters, and the best-fit spectral energy distribution (SED) of
the source are shown in Fig. \ref{fig:lepton}. The best fit model
parameters correspond to the peak of the 1-D probability distributions.
The spectral parameters in the leptonic scenario are well constrained
except for $n_{\rm ISM}$, whose $2\sigma$ upper limit is well determined.
In this calculation we set $T_e=10^7$ K. The only parameter sensitive to
$T_e$ is $n_{\rm ISM}$ in the leptonic scenario. The $2\sigma$ upper limit
of $n_{\rm ISM}$ is $0.007$ cm$^{-3}$ for $T_e=10^7$ K. The dotted line
in the right panel of Fig. \ref{fig:lepton} indicates the thermal
emission for these parameters. Since the ISM density will be more
essential for the discussion of the hybrid and hadronic models, we will
discuss the $T_e$ dependence of these results in detail in the subsection 2.4.

For the 2-D confidence regions of the parameters, we only show combinations
with relatively large correlation. There are very weak correlations among
$\alpha_e$, $E_c^e$, $W_e$ and $B$. The weak correlation between $\alpha_e$
and $W_e$ is mostly due to the facts that electrons near the high-energy
cutoff are well-constrained by observations and low-energy electrons
contribute the most to $W_e$ for $\alpha_e>2$. The correlation between
$\delta_e$ and $E_c^e$ is caused by the well observed spectral shape in
hard X-rays and TeV $\gamma$-rays. The combination of the X-ray and
$\gamma$-ray data helps to determine the model parameters. In the following
we will see that X-ray data alone lead to poorly constrained and highly
correlated parameters. This is an example to show the importance of global
fit to the multi-wavelength data.

The parameters and $\chi^2$ values of the best fit model are compiled in
Tables \ref{table:best} and \ref{table:chi2}, respectively.
Since $n_{\rm ISM}$ is not well constrained, its $2\sigma$ upper limit
instead of the best fit value is listed in Table \ref{table:best}.
The best-fit parameters are consistent with previous studies
\citep{2006A&A...449..223A,2008ApJ...683L.163L,2008ApJ...685..988T} and 
can be readily accommodated with typical SNRs \citep{2010ApJ...712..287E}.
The overall $\chi^2$ of the fit is relatively large with the reduced
$\chi^2$ $\sim 466.9/232=2.01$. Such a high value of $\chi^2$ shows that 
systematic errors dominate. This is not surprising given the relatively 
complex structure of the SNR, uncertainties related to the particle 
acceleration process, and our rather simple prescription of the emission 
model. The systematic errors actually dominate in all emission models,
which is typical for modeling of astrophysical observations of relatively 
complex phenomena. The X-ray and TeV $\gamma$-ray data contribute the 
most to the overall $\chi^2$. Especially for the TeV $\gamma$-ray data, 
the $\chi^2$ value is $149$ for $27$ data points, corresponding to an 
average residuals about $2.3\sigma$.
That is to say this simple leptonic model actually can not fit the $HESS$
data well. This is a well-known result in previous studies \citep[e.g.,][]
{2006A&A...449..223A,2008ApJ...685..988T,2009MNRAS.392..240M,
2009MNRAS.392..925F}. In \cite{2008ApJ...683L.163L} the authors proposed
a stochastic acceleration model to generate the electron spectrum with
sub-exponential cutoff ($\delta_e=0.5$) to better fit the $HESS$ data.
However, in such a case the fit to X-ray data becomes worse.
The X-ray data actually favors super-exponential cutoff instead (with
$\delta_e=1.2$ in this purely leptonic fit). The fit may be improved in
some detailed leptonic models, as shown in \cite{2010A&A...517L...4F}, 
though systematic errors still dominate.

\begin{table}[!htb]
\centering
\caption{Fitting parameters for $T_e=10^7$ K. Errors are $1\sigma$
statistical uncertainties; limits correspond to $2\sigma$ confidence level.}
{\scriptsize
\begin{tabular}{cccccccccc}
\hline \hline   & $\alpha_e$ & $E_c^e$ & $W_e$ & $\delta_e$ & $B$ & $n_{\rm ISM}$ & $\alpha_p$ & $E_c^p$ & $W_p$\\
  & & (TeV) & ($10^{47}$erg) & & ($\mu$G) & ($10^{-2}$cm$^{-3}$) & & (TeV) & ($10^{52}$erg)\\
\hline
  leptonic & $2.15^{+0.01}_{-0.01}$ & $51.3^{+2.3}_{-2.2}$ & $5.5^{+0.3}_{-0.3}$ & $1.21^{+0.04}_{-0.04}$ & $11.6^{+0.1}_{-0.1}$ & $<0.7$ & --- & --- & --- \\
  hadronic & $1.64^{+0.09}_{-0.08}$ & $14.5^{+4.8}_{-3.9}$ & $0.05^{+0.05}_{-0.02}$ & $2.1^{+0.2}_{-0.2}$ & $428.2^{+233.9}_{-159.6}$ & $<1.1$ & $1.58^{+0.06}_{-0.06}$ & $53.7^{+7.1}_{-6.2}$ & $>1.6$ \\
  hadronic$^*$ & $1.58^{+0.05}_{-0.05}$ & $12.3^{+2.1}_{-1.8}$ & $0.03^{+0.01}_{-0.01}$ & $1.9^{+0.1}_{-0.1}$ & $596.8^{+173.0}_{-129.0}$ & $<1.2$ & --- & $54.7^{+6.0}_{-5.7}$ & $>1.4$ \\
  hybrid   & $2.14^{+0.01}_{-0.01}$ & $50.7^{+2.1}_{-2.0}$ & $4.6^{+0.3}_{-0.3}$ & $1.23^{+0.04}_{-0.04}$ & $12.0^{+0.2}_{-0.2}$ & $<0.9$ & --- & --- & $>1.0$ \\
  \hline
  \hline
\end{tabular}}
\label{table:best}
\end{table}

\begin{table}[!htb]
\centering
\caption{Best-fit $\chi^2$ values for each set of data and the total reduced
one, for $T_e=10^7$ K.}
\begin{tabular}{cccccc}
\hline \hline   & radio & X-ray & GeV  & TeV & reduced \\
\hline
  leptonic     & $2.07$  & $312.0$ & $3.4$  & $149.4$ & $466.9/232$ \\
  hadronic     & $0.13$  & $291.7$ & $1.9$  & $40.6$  & $334.4/229$ \\
  hadronic$^*$ & $0.10$  & $291.9$ & $1.9$  & $40.5$  & $334.4/230$ \\
  hybrid       & $0.02$  & $305.4$ & $20.0$ & $109.3$ & $434.7/231$ \\
  \hline
  \hline
\end{tabular}
\label{table:chi2}
\end{table}

\subsection{Hadronic scenario}

In this subsection we discuss the model with a predominantly hadronic origin
of the $\gamma$-rays. The spectrum of the accelerated protons is assumed
to be $F_p(E)\propto E^{-\alpha_p}\exp[-(E/E_c^p)^{\delta_p}]$ with
$\delta_p=1$, which gives acceptable fit to the TeV data. The normalization
is fixed using the total kinetic energy of protons with the energy $E>1$ GeV.
For the hadronic $\gamma$-ray production we adopt the parameterization
of \cite{2006ApJ...647..692K}. With the additional three parameters,
$\alpha_p,\,E_c^p,\,W_p$, we have 9 parameters in total. Considering the
synchrotron cooling\footnote{The IC cooling is negligible compared with
the synchrotron cooling for magnetic fields greater than $10 \mu$G.}
of high energy electrons, we also introduce a spectral break to the overall
electron distribution\footnote{The break energy in the leptonic (and the
following hybrid) model is very high due to the weak magnetic field, and
the electron spectrum is identical to a single power law.}.
The break energy, at which the synchrotron cooling
time is equal to the lifetime of the remnant, is determined with
$E_{\rm br}\approx 7.8\times10^{6}(B/{\rm \mu G})^{-2}(T_{\rm life}/{\rm
1600\,yr})^{-1}{\rm GeV}$ \citep{2008ApJ...685..988T}. For $E_e<E_{\rm br}$
the power-law index is $\alpha_e$, and for $E_e>E_{\rm br}$ the power-law
index is $\alpha_e+1$. Since $T_{\rm life}$ is taken as 1600 years,
$E_{\rm br}$ is not a free parameter.

\begin{figure}[!htb]
\centering
\includegraphics[width=0.32\columnwidth]{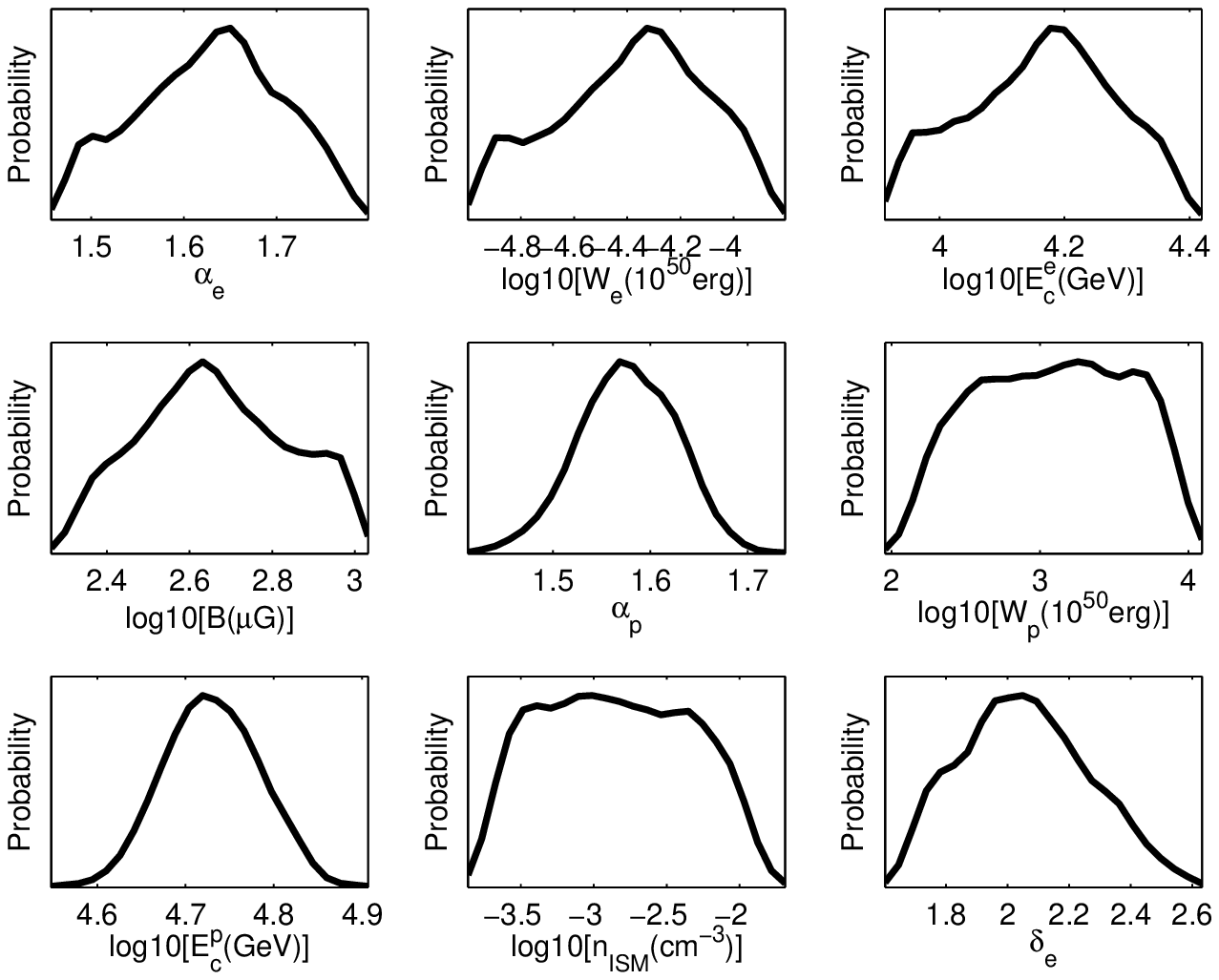}
\includegraphics[width=0.32\columnwidth]{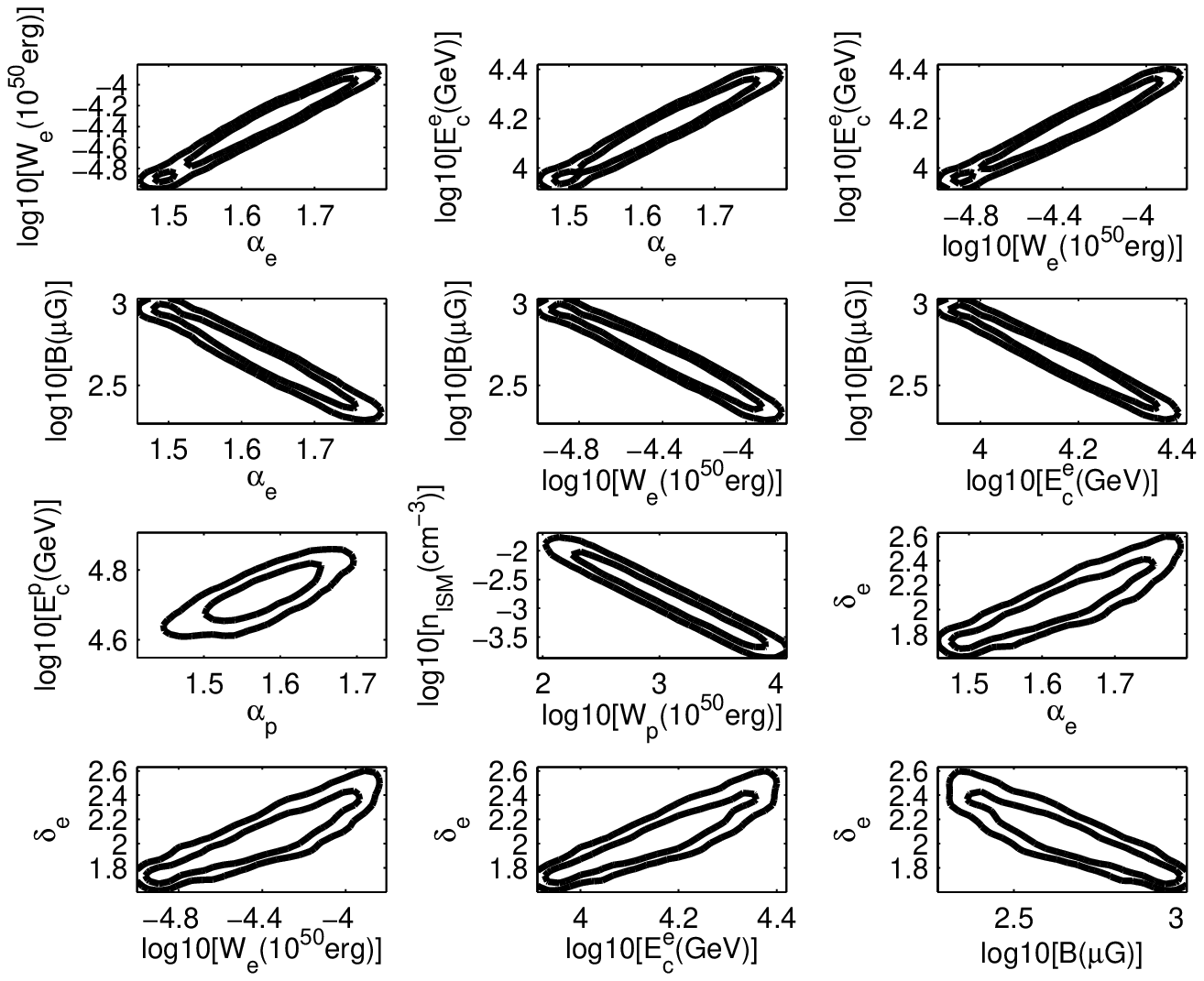}
\includegraphics[width=0.32\columnwidth]{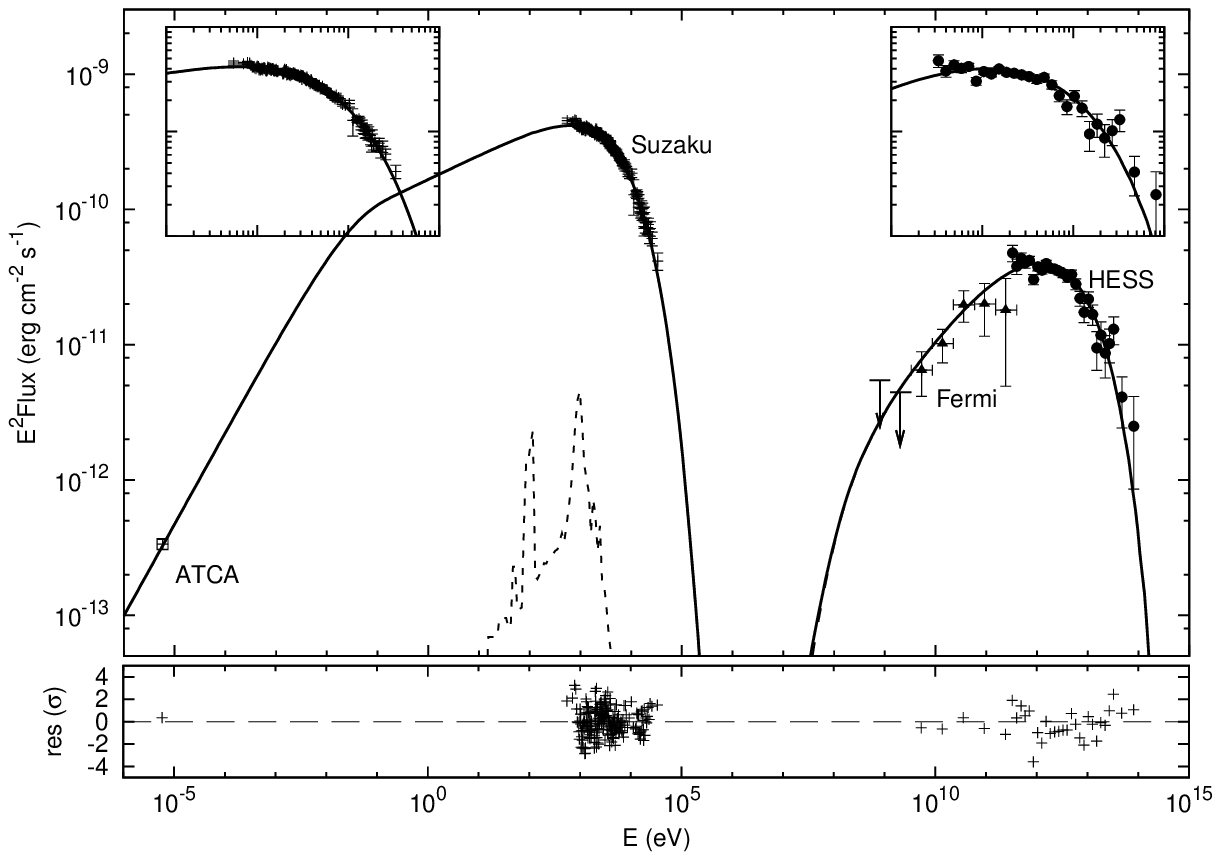}
\caption{Same as Fig. \ref{fig:lepton} but for the hadronic scenario.
The dashed line indicates the thermal emission for $T_e=10^7$ K and
$n_{\rm ISM} = 0.011$ cm$^{-3}$.
}
\label{fig:hadron}
\end{figure}

The 1-D probability distributions and 2-D confidence contours of the
model parameters, and the SED of the best-fit model are shown in Fig.
\ref{fig:hadron}. The parameters and $\chi^2$ values of the best fit are
listed in Tables \ref{table:best} and \ref{table:chi2}, respectively.
Still we adopt $T_e=10^7$ K in this calculation. Compared with the
leptonic model, a much stronger magnetic field is inferred, which is
consistent with previous studies \citep{2006A&A...451..981B} and the
interpretation of the observed X-ray surface brightness fluctuations
as synchrotron cooling of high energy electrons \citep{2007Natur.449..576U}.
The 1-D probability distribution of most model parameters do not converge 
very well with multiple peaks except for those for $\alpha_p$, and $E_c^p$, 
which are constrained by the $\gamma$-ray data directly, independent of 
other parameters in the hadronic scenario.

Due to the lack of constraint from the $\gamma$-ray data, the parameters
related to electron emission can not be well determined with relatively
large $1\sigma$ errors shown in Table \ref{table:best}, and there are
strong correlations in the 2-D confidence contours, as shown in the
middle panel of Fig. \ref{fig:hadron}. The correlations among $W_e$,
$E_c^e$, and $B$ are due to the facts that the synchrotron emissivity
$\epsilon\propto W_eB^2E_c^{e2}$, and the high energy cutoff of synchrotron
emission $\nu_c\propto B E_c^{e2}$. $\nu_c$ is well constrained by
X-ray observations, which leads to the anti-correlation between $B$
and $E_c^{e}$. The anti-correlation between $W_e$ and $B$ can be attributed
to the measured luminosity of the synchrotron emission. The correlation
between $W_e$ and $E_c^{e}$ results from these two anti-correlations.
The strong anti-correlation between $B$ and $\alpha_e$ is due to the
radio to X-ray spectral shape, which, in combination with the
anti-correlations between $B$ and $E_c^e$, and $B$ and $W_e$, leads to
the correlations between $\alpha_e$ and $E_c^e$, and $\alpha_e$ and
$W_e$, respectively. The correlations related to $\delta_e$ can be
attributed to the hard X-ray spectrum. The weak correlation between
$\alpha_p$ and $E_c^p$ is due to the well-measured high-energy cutoff
of the TeV emission. The strong anti-correlation between $n_{\rm ISM}$
and $W_p$ is due to the fact that the product of $n_{\rm ISM}$ and $W_p$
determines the hadronic component of $\gamma$-ray emission. Therefore we
can get a lower limit of $W_p$ according to the upper limit of $n_{\rm ISM}$.
The $2\sigma$ upper limit of $n_{\rm ISM}$ is $0.01$ cm$^{-3}$ for
$T_e=10^7$ K, corresponding to a lower limit of $1.6\times 10^{52}$ erg
for $W_p$. The dependence of these results on $T_e$ will be discussed in
subsection 2.4. The constraint on $W_p$ requires a total energy of CR
protons much higher than typical energy output of a supernova explosion,
say $10^{51}$ erg \citep{2010ApJ...712..287E,2010ApJ...708..965Z}.

Table \ref{table:chi2} shows clearly that the fits to GeV and TeV
$\gamma$-ray data are significantly improved in the hadronic model.
The average residual for the $HESS$ data becomes $\sim 1.14\sigma$.
The fit to the X-ray data also improves a bit due to a larger value of
$\delta_e$. Note, however, the reduced $\chi^2$ of the
global fit is about $1.46$, which is still too large to be attributed to 
pure statistical errors. There are some systematical effects in either 
the data or the model. In particular, contributions to the $\chi^2$ are 
dominated by the X-ray data in all these models, as shown in Table 2.
It suggests that our simple one-zone synchrotron emission model does not
give a sufficient description of the spatially integrated emission spectrum.
Compared with the leptonic model, most reduction in the value of the $\chi^2$ 
of the hadronic model comes from improved fit to the $\gamma$-ray data. 

In the hadronic model the magnetic field is large, which suppresses the
IC contribution to the $\gamma$-rays from energetic electrons and makes
hadronic contribution to the $\gamma$-ray dominant. The strong magnetic
field of the best fit model implies very efficient energy loss near the
cutoff energy of electrons, which not only introduces a spectral break
in the overall electron distribution but can also render the high energy
cutoff sharper with $\delta_e=2.1$ \citep{2010MNRAS.402.2807B}.
It is interesting to note that the spectral index for electrons $\alpha_e$
is consistent with that for protons $\alpha_p$, which is expected if the
acceleration of these high-energy particles is due to the same physical
process. The difference in their high-energy cutoffs can be attributed
to the difference in the energy loss rate of protons and electrons near
the cutoff energy. Therefore $E_c^e$ should
not be compared to $E_e^p$ directly.

\begin{figure}[!htb]
\centering
\includegraphics[width=0.32\columnwidth]{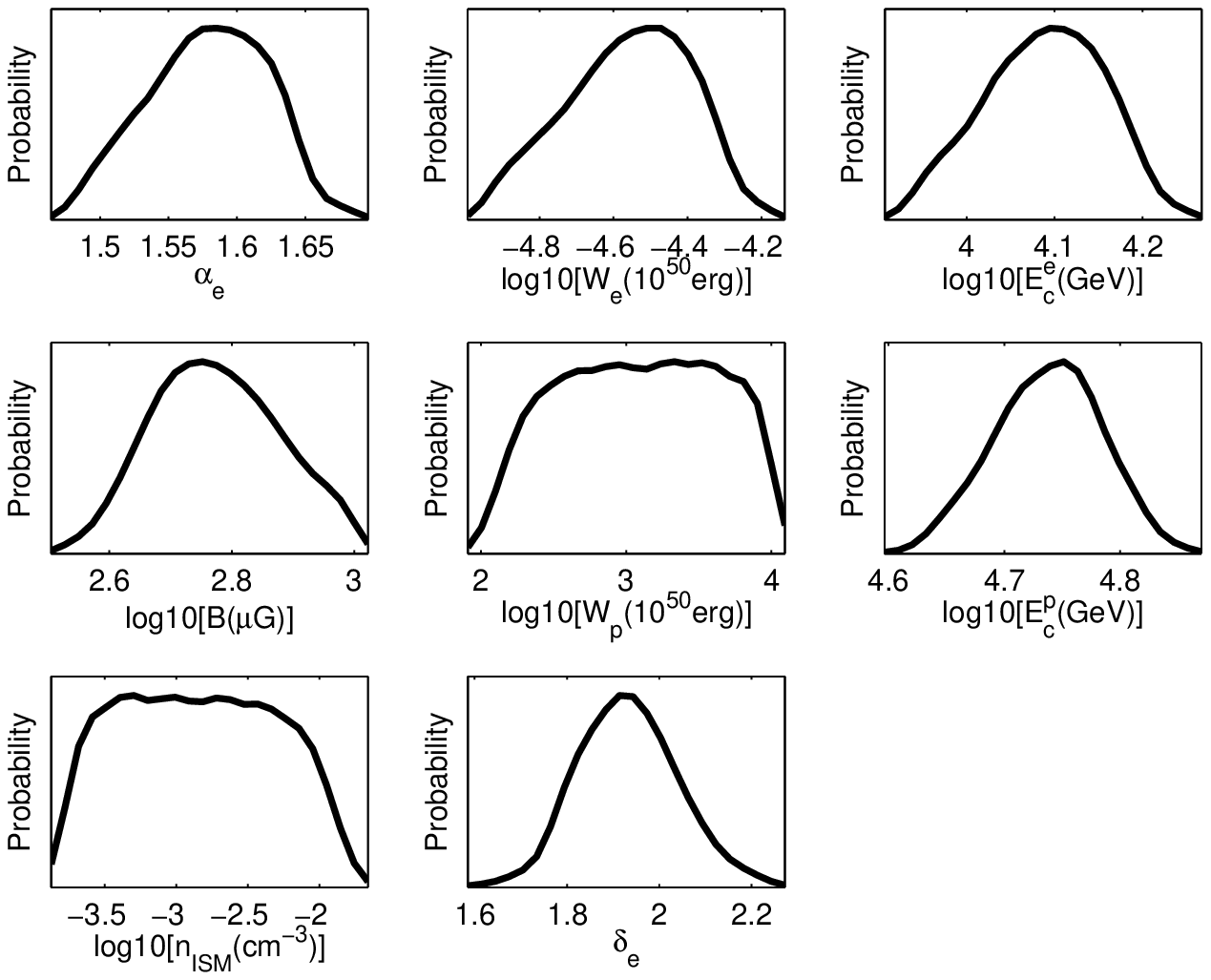}
\includegraphics[width=0.32\columnwidth]{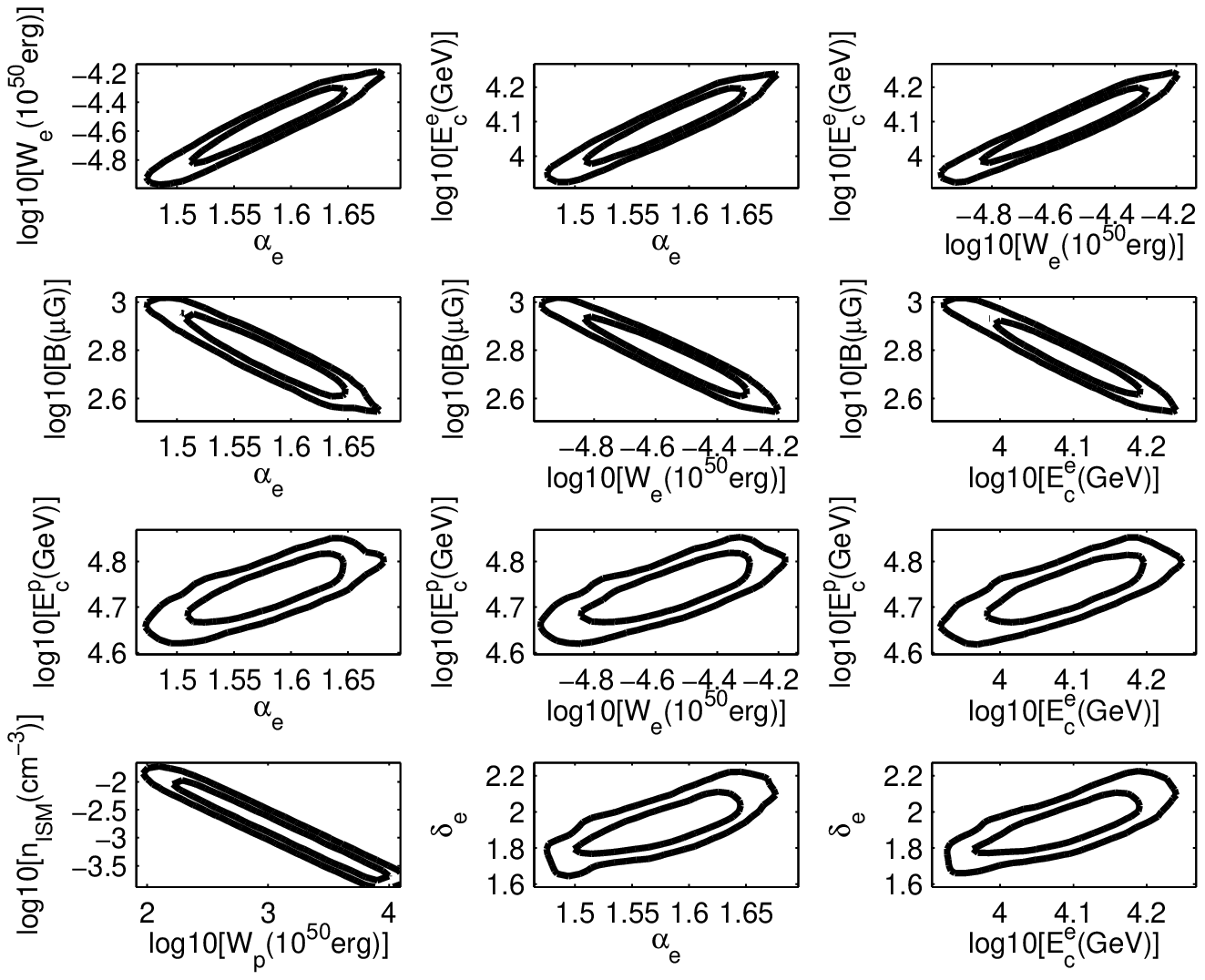}
\includegraphics[width=0.32\columnwidth]{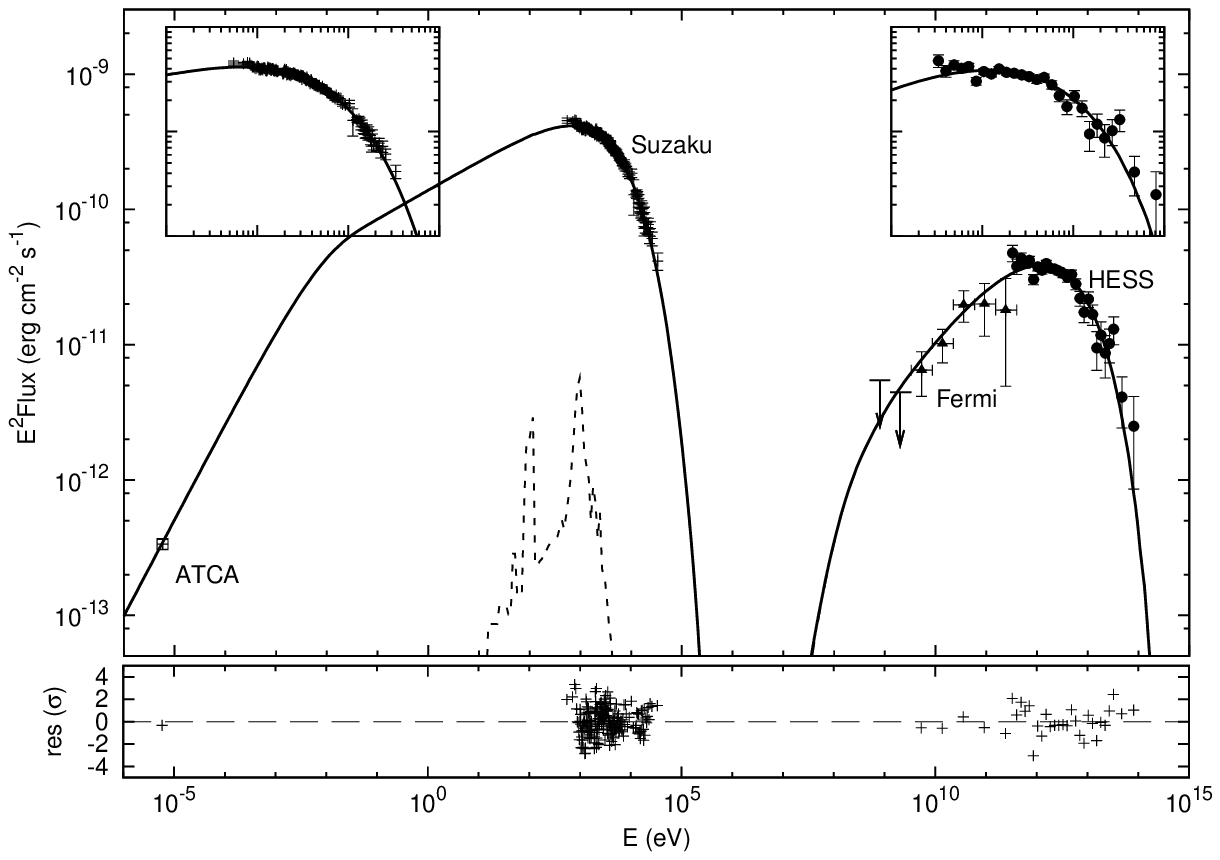}
\caption{Same as Fig. \ref{fig:hadron} but for the hadronic scenario
with the extra requirement of $\alpha_e=\alpha_p$.
}
\label{fig:hadron2}
\end{figure}

To reduce uncertainties of the model
parameters in the hadronic scenario, we also consider the
hadronic model with the constraint that $\alpha_e=\alpha_p$. The 1-D
probability distribution and the 2-D confidence contours, and the best
fit SED for such a model are shown in Fig. \ref{fig:hadron2}. Compared
with Fig. \ref{fig:hadron}, the probability distribution of the model
parameters are better converged and the correlations in the 2-D confidence
contours are weakened. The acceptable model parameter space is reduced
significantly. The fitting results are also compiled in Tables
\ref{table:best} and \ref{table:chi2} with a star mark. The value of the
best fit model parameters agree with those of the hadronic model and the
$1\sigma$ error of parameters related the electron emission are reduced
significantly. The values of the $\chi^2$ are essentially the same as 
the hadronic model. 

The value of the spectral index for the best fit model is always much
less than 2, a result difficult to accommodate with the diffusive shock
model. This in combination with the low limit on $W_p$ poses one of the
most serious challenges to the hadronic scenario in the context of
diffusive shock acceleration of SNRs. We also note that the best fit 
electron spectral index $\alpha_e$ is harder for the hadronic model 
than that for the leptonic model. Although the current radio data does 
not give a good constraint on synchrotron spectral index, a better 
measurement of the synchrotron spectrum in the future will be helpful 
in distinguishing these two scenarios for the TeV emission.

\subsection{Hybrid scenario}

The results above agree with previous studies
\citep{2008ApJ...685..988T,2006A&A...449..223A}. They demonstrate clearly
the strengths and problems associated with the leptonic and hadronic
scenarios. The leptonic model has fewer parameters, most of which are
well constrained with the MCMC method by fitting the SED. The fact that
it can give reasonable good fits to the overall SED with reasonable values of
the parameters may be considered as evidence for such a scenario
\citep{2010MNRAS.406.1337F}. On the other hand, the relatively high values
of $\chi^2$, especially for the $\gamma$-ray data, suggest that the model
may not be complete. It has been shown that the TeV emission from SNR 
RX J1713.7-3946 may have significant
contributions from energetic protons as well \citep{2010ApJ...708..965Z,
2008JCAP...01..018K}. However, in the most general case, at least three
more parameters need to be introduced to characterize the distribution
of accelerated protons, which leads to strong degeneracy of the model
parameter space. And the challenges to the hadronic scenario, namely
hard spectra of accelerated particles and the lower limit on $W_p$, does
not appear to depend on this degeneracy in the regime of parameter space
explored above.

In the subsection 2.2, we demonstrate that the acceptable
model parameter space may be reduced significantly by considering some
physically motivated constraint on the accelerated particle distributions.
We have argued that the spectral indices of accelerated electrons and
protons should be comparable in the relativistic energy regime where 
the energy loss can be ignored. In the hadronic scenario, due to the 
presence of strong magnetic field, there is a spectral break in the 
electron distribution and the high energy cutoffs of electrons and protons 
do not need to be the same. However, in the leptonic scenario, the magnetic 
field is so weak that the energy loss does not affect the distribution 
of both electrons and protons. For these high energy relativistic 
electrons and protons, their gyro-radius only depends their energy and 
the magnetic field. We would expect that mechanisms of charged particle 
acceleration will lead to identical particle distributions except their 
normalization, which is determined by different injection processes at 
low energies \citep{PL04}.
To reduce the number of model parameters, one may therefore consider
the hybrid scenario where $\alpha_e=\alpha_p$, $\delta_e=\delta_p$, and
$E_c^e=E_c^p$. As we will show below, this leads to a hybrid explanation
to the high energy $\gamma$-ray data. The total number of free parameters
in the hybrid model is now only 7.

\begin{figure}[!htb]
\centering
\includegraphics[width=0.32\columnwidth]{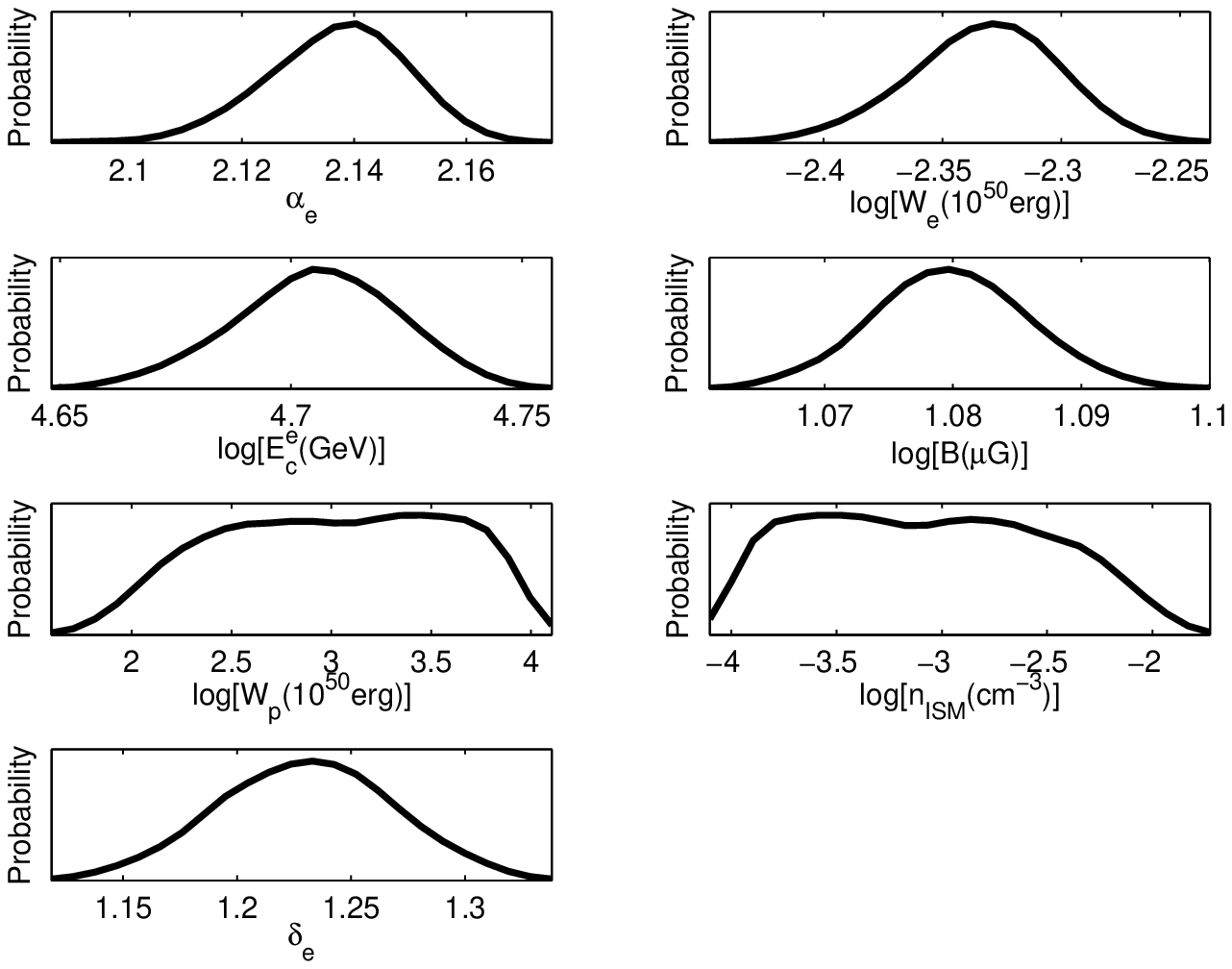}
\includegraphics[width=0.32\columnwidth]{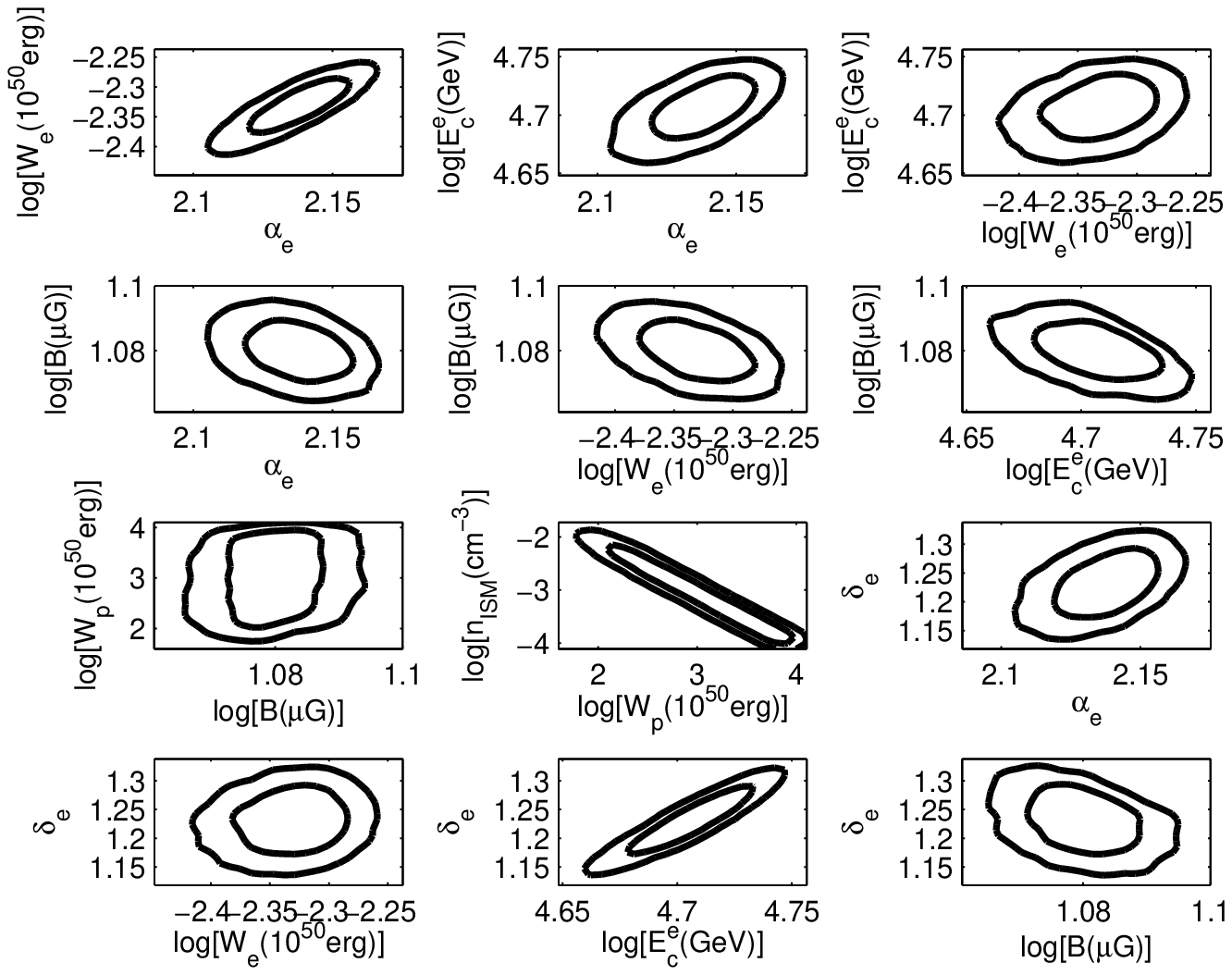}
\includegraphics[width=0.32\columnwidth]{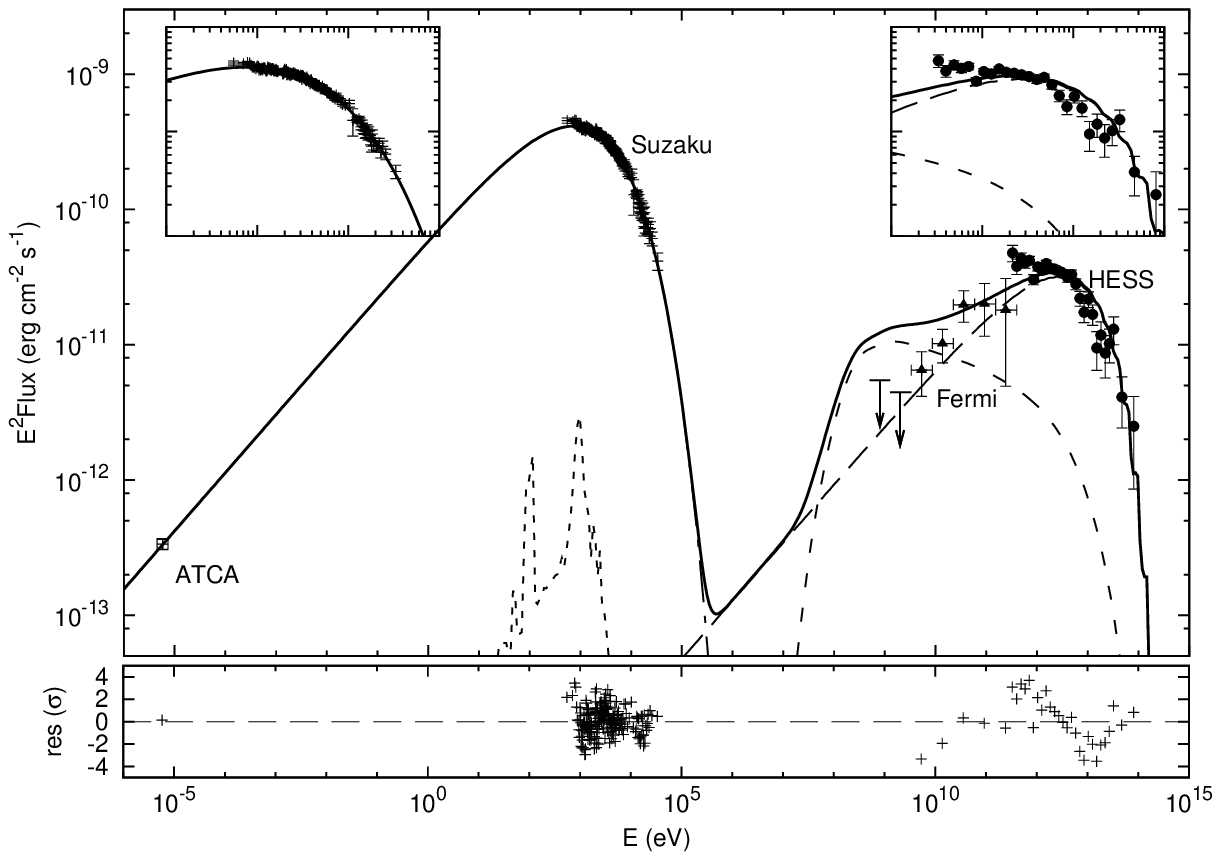}
\caption{Same as Fig. \ref{fig:lepton} but for the hybrid scenario.
The density for the thermal emission indicated by the dashed line is
$0.009$ cm$^{-3}$.
}
\label{fig:hybrid}
\end{figure}

The model parameter distributions and best-fit SED are shown in Fig.
\ref{fig:hybrid}. The background electron temperature is still adopted
as $10^7$ K. The best fit results and $\chi^2$ values are also listed in
Tables \ref{table:best} and \ref{table:chi2}. We see that the parameters
of the electron component do not change significantly compared with the
leptonic scenario primarily due to the dominance of TeV emission by
relativistic electrons through the IC process. The hadronic component
dominates the GeV $\gamma$-ray emission. Note that since $E_c^p=E_c^e$,
the $\gamma$-ray spectrum of the hadronic component cuts off at a lower
energy than the leptonic component. Neutral pion decays into two
$\gamma$-ray photos. The cutoff energy of the $\gamma$-ray spectrum is
at least a factor of 2 lower than the cutoff energy of the corresponding
proton distribution. For the IC emission, the cutoff energy of the
$\gamma$-ray spectrum can be the same as the corresponding electron
distribution. The fit to TeV data shows improvement compared with the
leptonic model, with average residual changing from $\sim 2.3\sigma$ to
about $2.0\sigma$. However, the hadronic component seems to overproduce
the GeV flux, resulting in an even larger $\chi^2_{\rm GeV}$ than the
leptonic scenario. The $\chi^2$ value for X-ray data does not change
significantly. The correlation of the model parameters are also similar to
the leptonic model. And the strong anti-correlation between $n_{\rm ISM}$
and $W_p$ is still due to the fact that the observed emission is
determined by the product of the two. Although the model has a weak
magnetic field and relatively soft distributions of accelerated particles,
the $2\sigma$ lower limit of the proton energy of $1.0\times 10^{52}$
ergs is comparable to those of the hadronic models, which still challenges
the energetics of the SNR.

\subsection{Dependence of $T_e$}

The energy content of relativistic protons is poorly determined due to
the high uncertainty in $n_{\rm ISM}$. Perhaps the only observation
one can use to constrain $n_{\rm ISM}$ is the lack of thermal X-ray
emission from the remnant \citep{2004A&A...427..199C}. To derive a
robust constraint on $n_{\rm ISM}$, one however needs to consider the
heating of electrons and the ionization of ions in the background plasma,
both of which are not well understood though a preliminary attempt has been
taken to model these processes quantitatively \citep{2010ApJ...712..287E}.
Here we assume that electrons have reached ionization equilibrium with
the ions and so that the Raymond-Smith code can be used to calculate
the thermal emission.

\begin{figure}[!htb]
\centering
\includegraphics[width=0.32\columnwidth]{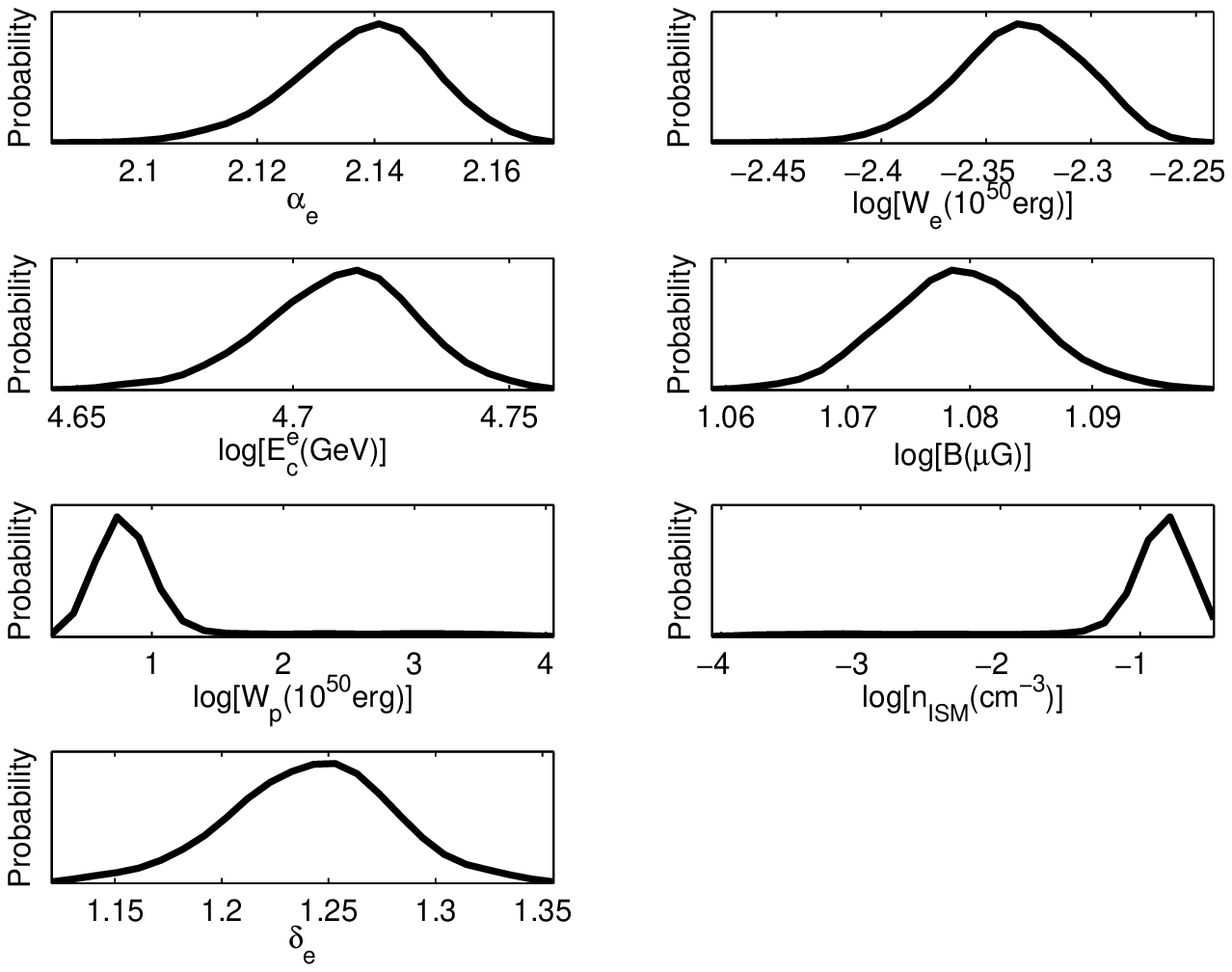}
\includegraphics[width=0.32\columnwidth]{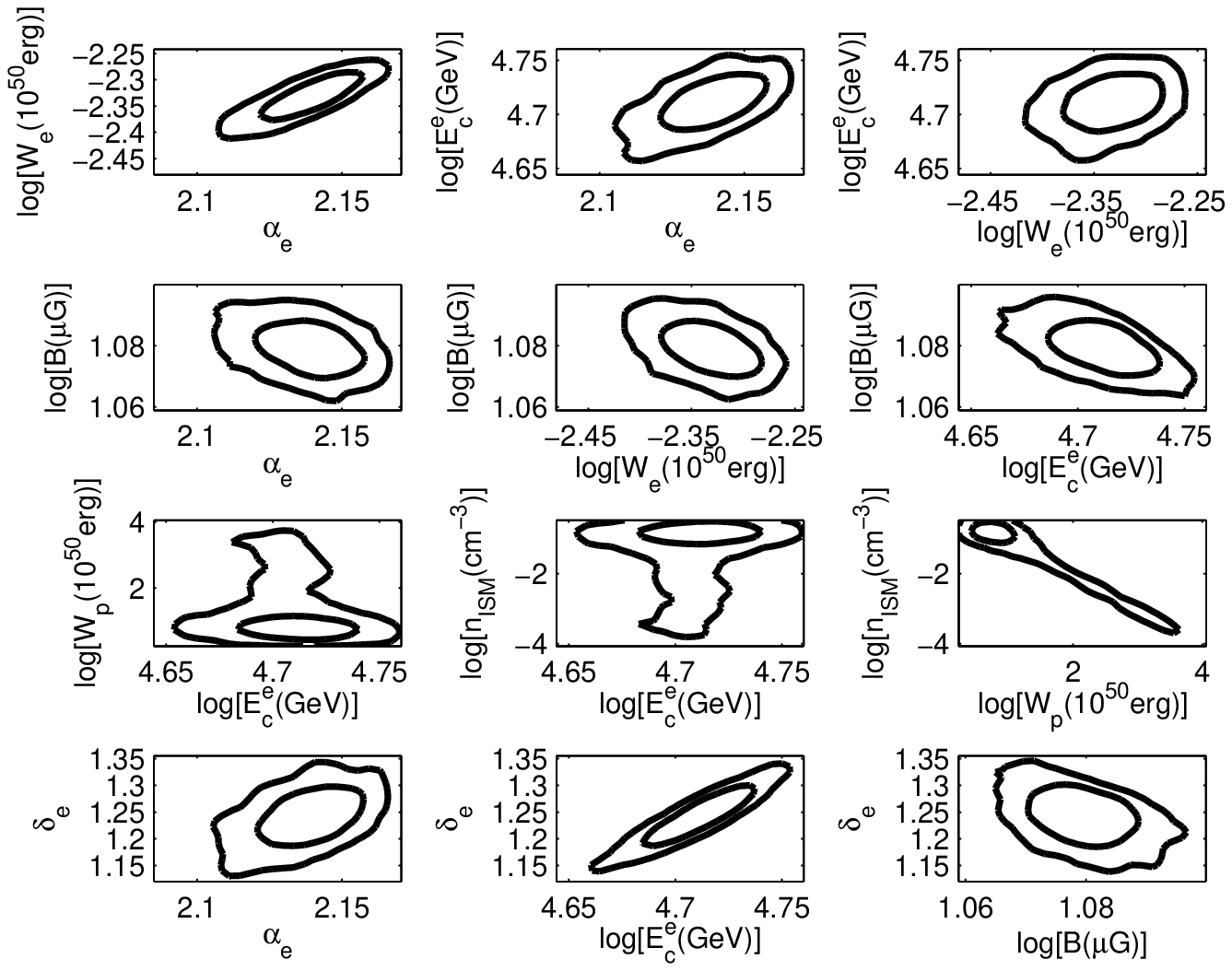}
\includegraphics[width=0.32\columnwidth]{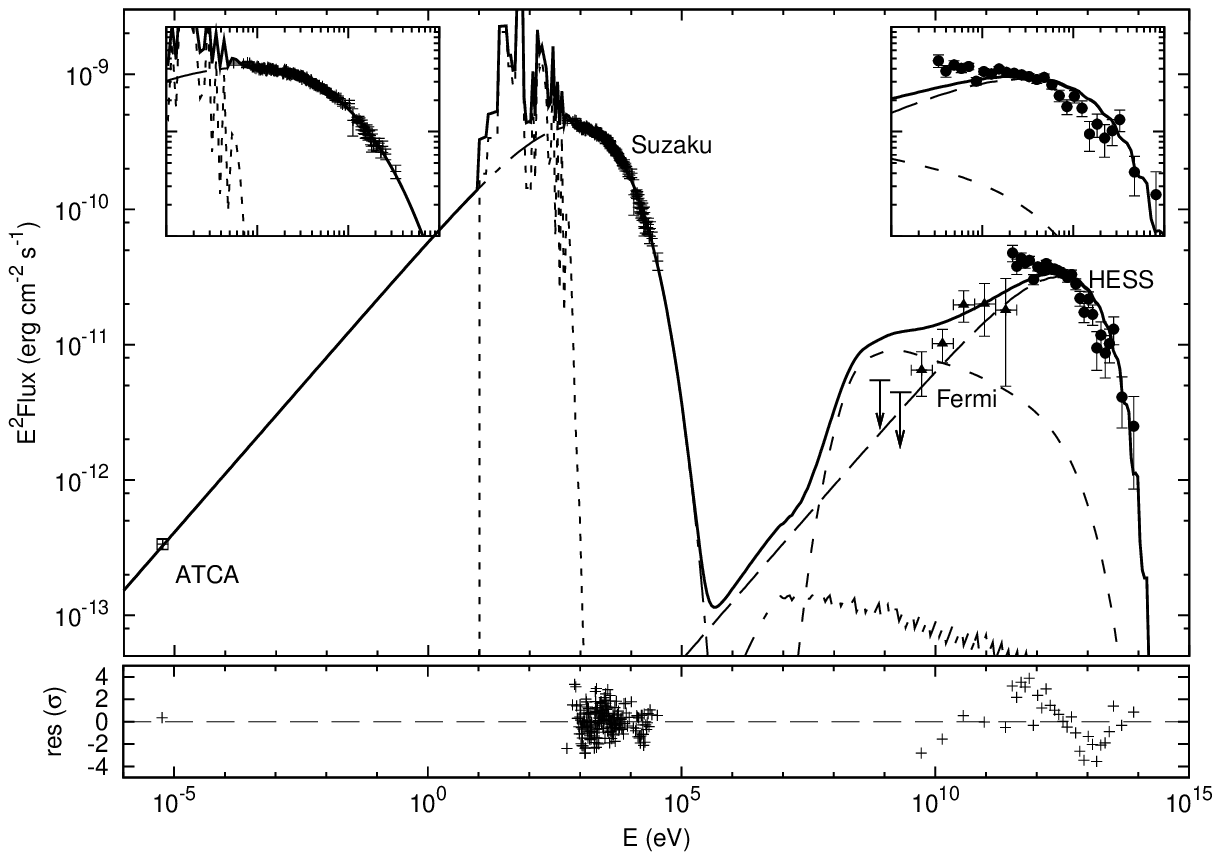}
\includegraphics[width=0.32\columnwidth]{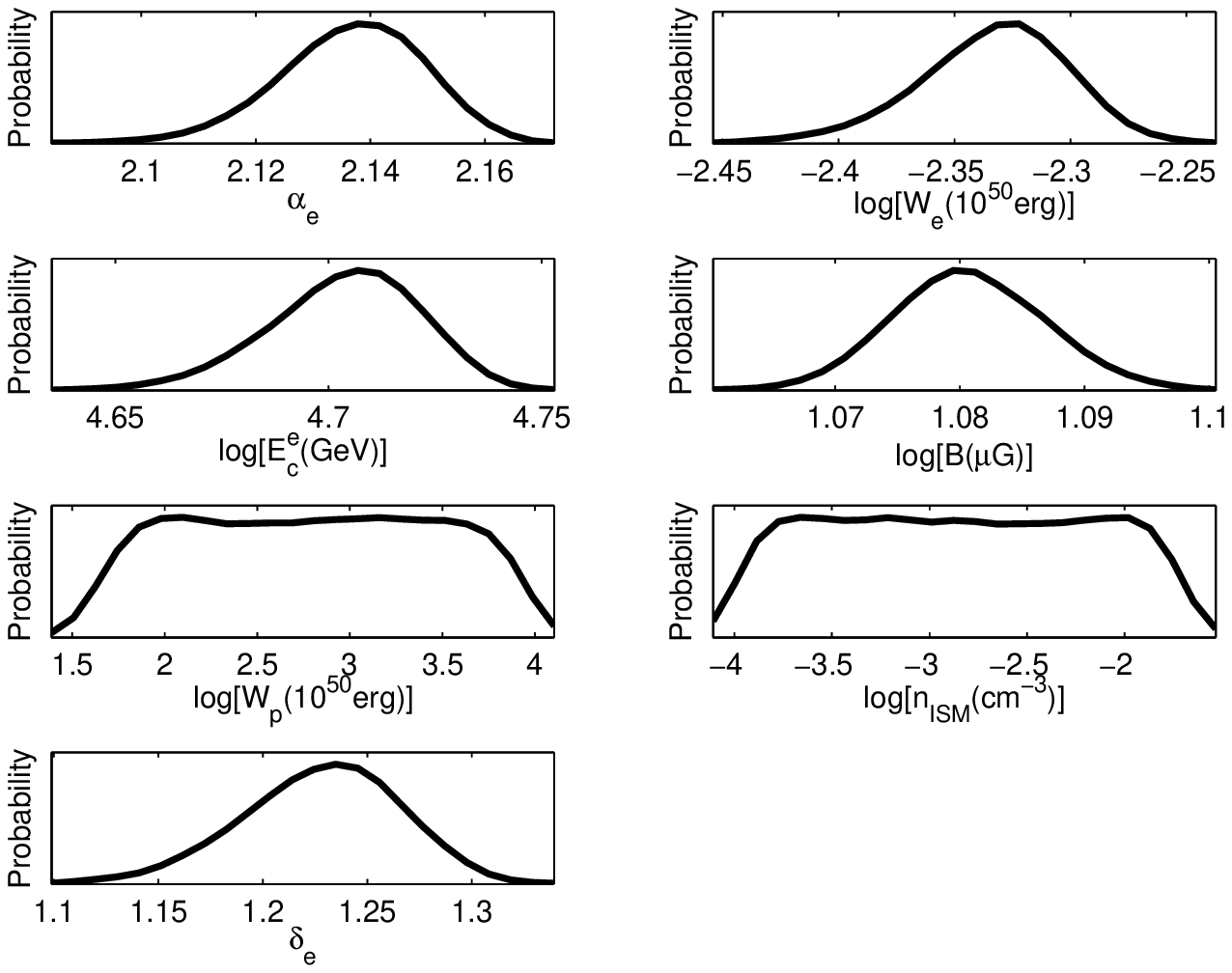}
\includegraphics[width=0.32\columnwidth]{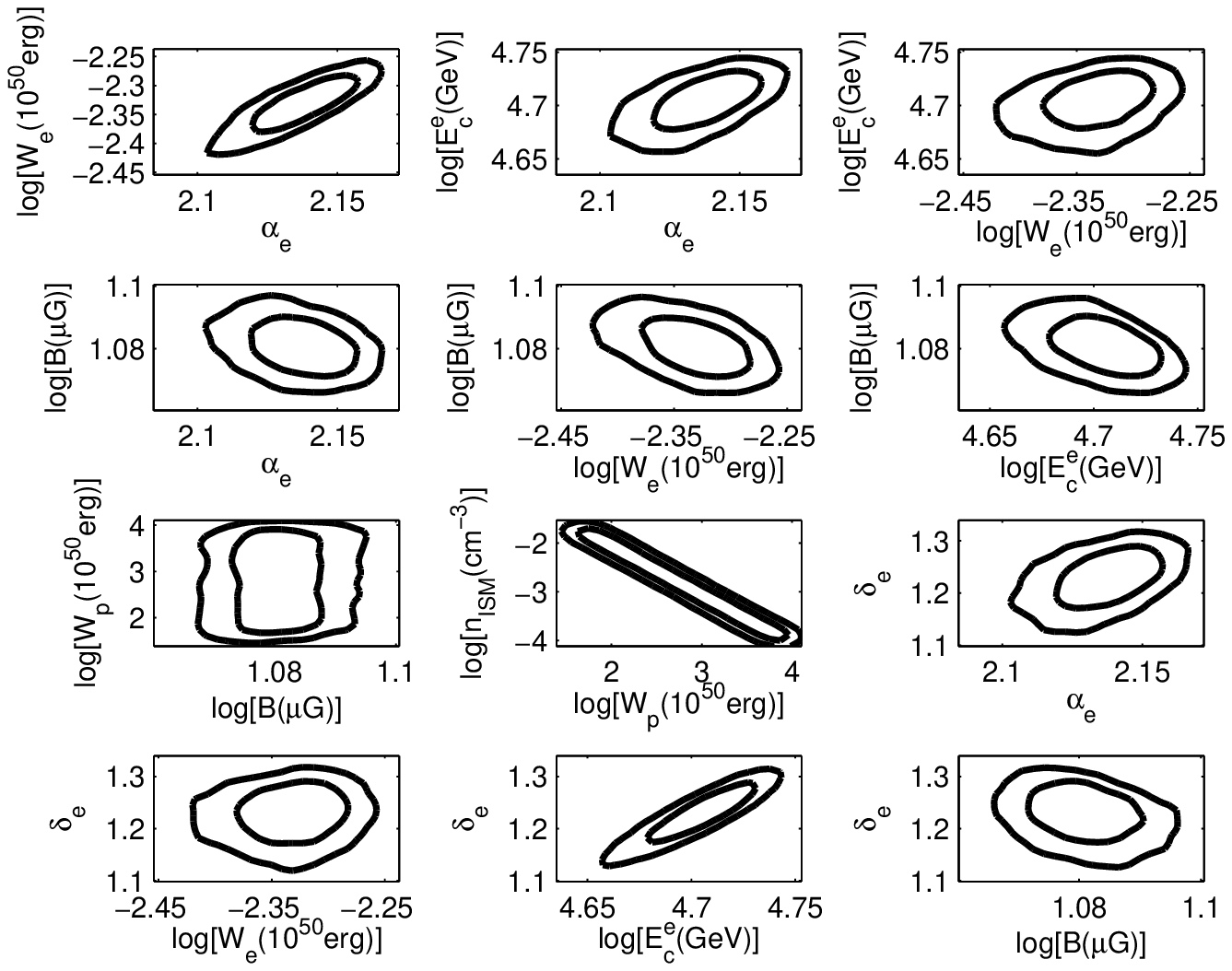}
\includegraphics[width=0.32\columnwidth]{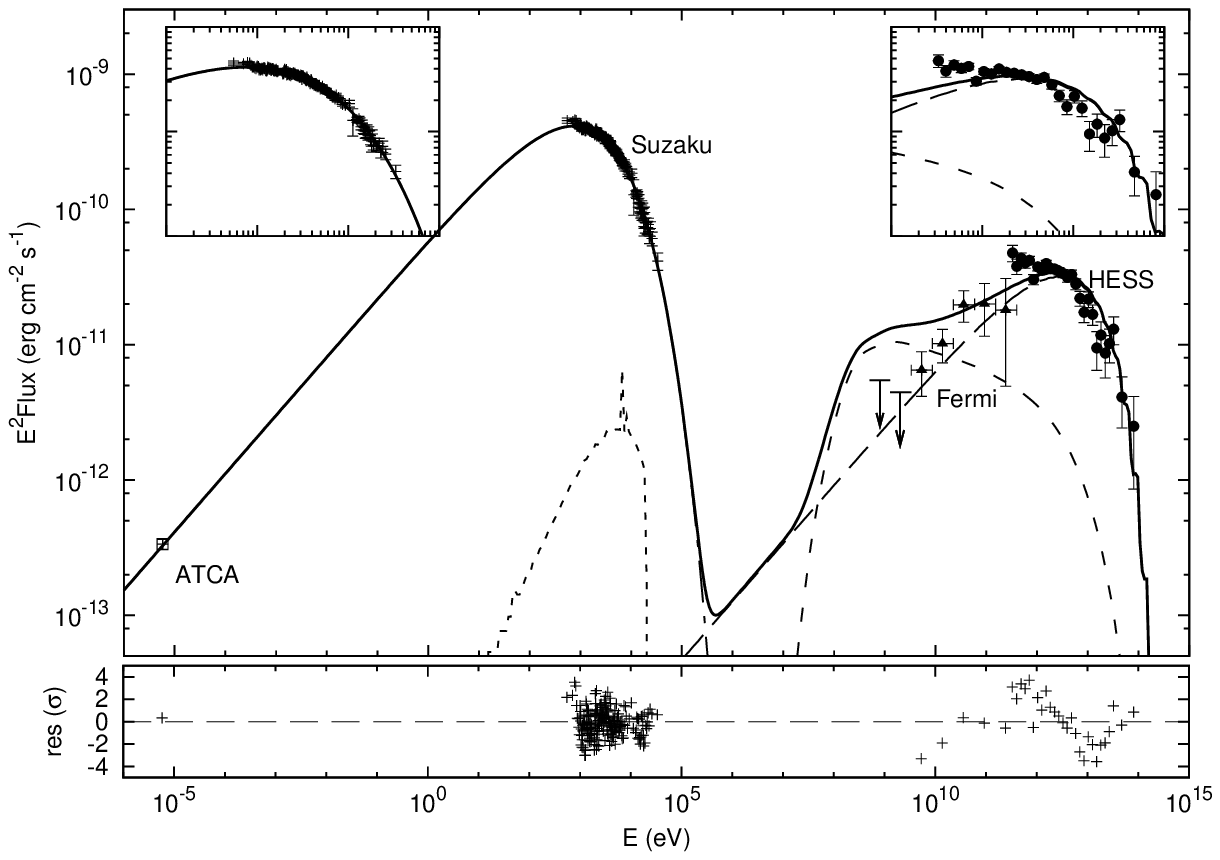}
\caption{Same as Fig. \ref{fig:hybrid} but for two other background electron
temperatures: $10^6$ (upper) and $10^8$ (lower) K. The corresponding density
$n_{\rm ISM}$ for the thermal emission is $0.2$ cm$^{-3}$ (upper) and $0.02$
cm$^{-3}$ (lower).
}
\label{fig:hybrid-tem}
\end{figure}

The results above do not differ significantly for different values of
$T_e$ except for the constraint on $n_{\rm ISM}$ and accordingly $W_p$.
In Fig. \ref{fig:hybrid-tem} we show the results for the hybrid scenario
with $T_e=10^6$ and $10^8$ K. For $T_e=10^6$ K most of the line emission
has energies lower than 0.5 keV, which is below the lower limit of the
$Suzaku$ data. However, the emission in the X-ray band is sufficient to
lead to well-constrained $n_{\rm ISM}\sim 0.2$ cm$^{-3}$ and $W_p\sim 
10^{51}$ ergs. The model also predicts strong emission below the X-ray 
range. A significant thermal component also helps to slightly improve 
the fit to the X-ray data. The $2\sigma$ upper limit of $n_{\rm ISM}$ for
$T_e=10^6$ K is $0.2$ cm$^{-3}$, which is much looser than that for
$T_e=10^7$ K. For $T_e=10^8$ K the $2\sigma$ upper limit of
$n_{\rm ISM}$ is $0.02$ cm$^{-3}$, which is also higher than $0.009$
cm$^{-3}$ for $T_e=10^7$ K. The $2\sigma$ lower limits of $W_p$ are
$3.5\times 10^{50}$, $1.0\times 10^{52}$, and $4.4\times 10^{51}$ ergs
for $T_e=10^6$, $10^7$, and $10^8$ K respectively.

\begin{figure}[!htb]
\centering
\includegraphics[width=0.45\columnwidth]{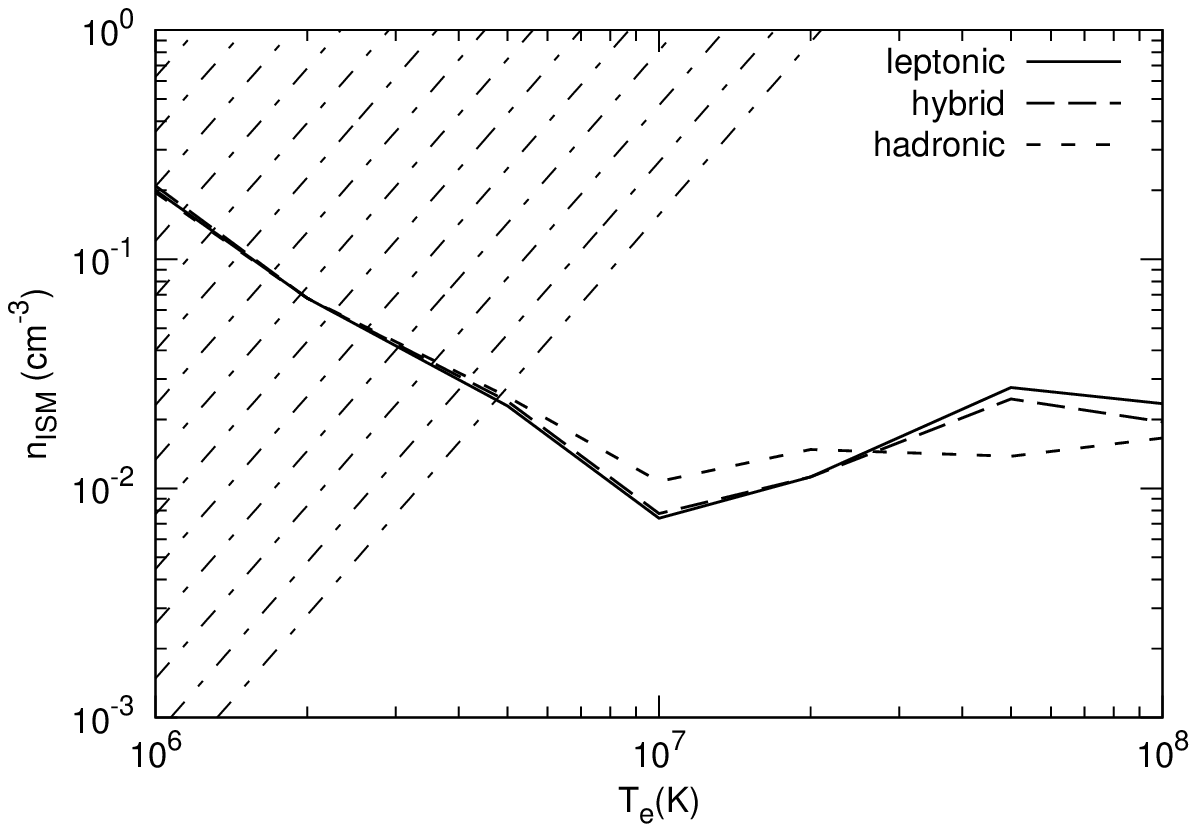}
\includegraphics[width=0.45\columnwidth]{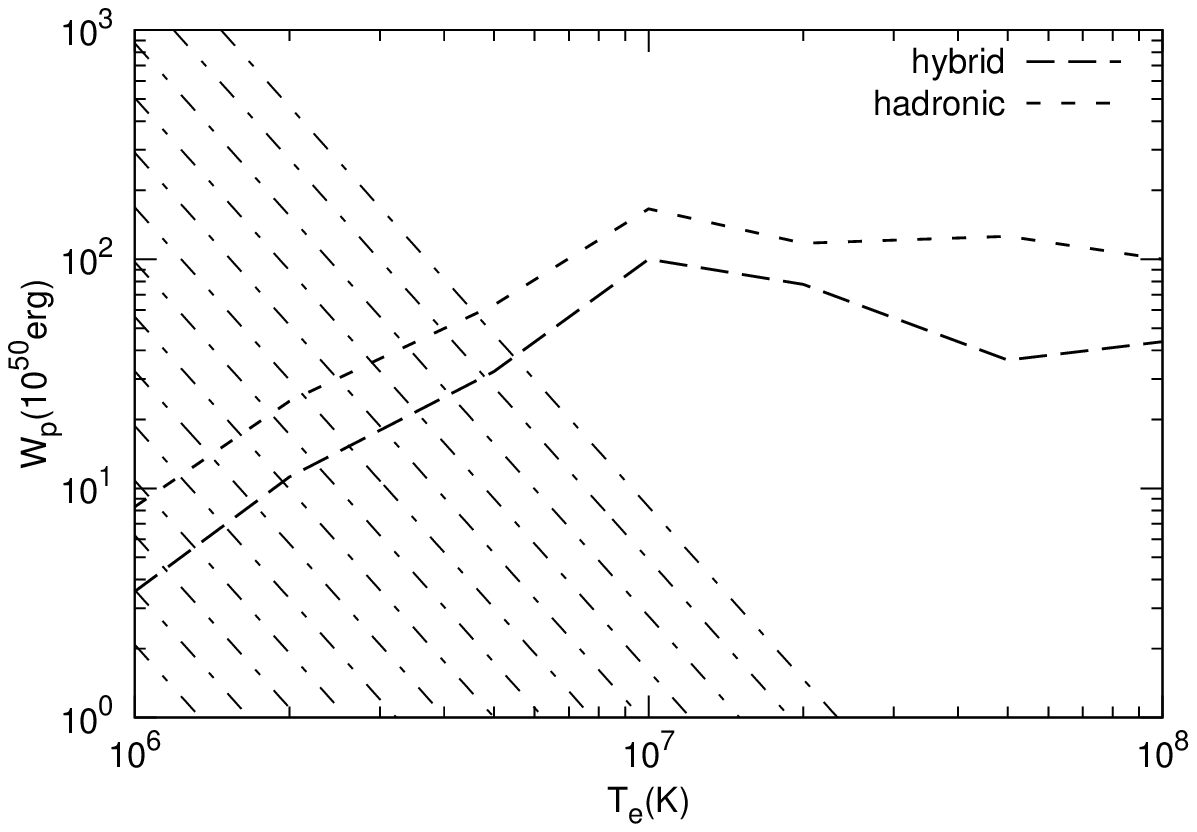}
\caption{The $2\sigma$ upper limits of $n_{\rm ISM}$ (left) and lower
limits of the energy content of non-thermal protons $W_p$ (right) as
functions of the temperature of background electrons $T_e$ for different
models. The shaded region can be excluded due to heating of electrons
by background protons through Coulomb collisions.
}
\label{fig:nism-wp}
\end{figure}

To demonstrate how the constraint on $n_{\rm ISM}$ and $W_p$ vary with
$T_e$, we repeat the MCMC calculation for a series of $T_e$. The $2\sigma$
upper limit of $n_{\rm SIM}$ and lower limit of $W_p$ are shown in Fig.
\ref{fig:nism-wp}. In the left panel the constraints on $n_{\rm ISM}$ of
the three scenarios are shown, while in the right panel the constraints
on $W_p$ are relevant for the hybrid and hadronic models. The constraints
on $n_{\rm ISM}$ are almost the same for the leptonic and hybrid models
since their parameters for the synchrotron emission are similar.
For the hadronic model the result is slightly different primarily due to
difference in the electron distribution. The lower limit on $W_p$ for
the hybrid scenario is a bit smaller than the hadronic model because in
the hybrid model the IC component from the lepton population has a
significant contribution to the $\gamma$-ray. The shaded region in the
left panel is excluded by considering the heating of electrons by ions
through Coulomb collisions with the scaling
relation of Eq. (1). In the right panel the shaded region is derived by
applying the relation $W_p/10^{50}\,{\rm erg}\sim 1.3\,{\rm cm}^{-3}/
n_{\rm ISM}$, which is the approximate relation required to reproduce
the high energy $\gamma$-ray emission for the hybrid and hadronic scenarios.
It can be seen that in general $W_p$ needs to be greater than $\sim 4\times
10^{51}$ erg for the hybrid model. For the hadronic scenario the requirement
of $W_p$ is even larger. Such a large value of proton energy seems to be
unacceptably high for a typical SNR \citep{2010ApJ...712..287E,
2010ApJ...708..965Z}. The 2$\sigma$ upper limit of $n_{\rm ISM}$ is
less than 0.03 cm$^{-3}$, which is consistent with that obtained by
\citet{2004A&A...427..199C} from {\it XMM-Newton} observations.

\section{Discussion and Conclusion}

Since the discovery of synchrotron X-ray emission from 
the forward shock of SN 1006 \citep{Ko95}, it has been established that 
shocks of SNRs can accelerate electrons to tens of TeV. The detection of 
TeV emission directly from a few shell type SNRs confirmed this conclusion 
\citep{2004ApJ...602..271L,aha05,aha08}. However, the nature of the 
TeV emission is still a matter of debate. X-ray observations also reveal 
strong thermal emission in the interior of a few young SNRs. Thermal 
X-ray emission from the forward shock, on the other hand, is usually weak 
and only detected in relatively old remnants, such as RCW 86 
\citep{2006ApJ...648L..33}. Since the observed TeV emission is well 
correlated with the forward shock, the lack of thermal X-ray emission 
implies low gas density and therefore high energy content of accelerated 
protons in the hadronic model for the TeV emission, which is difficult 
to accommodate with the SNR theories. The leptonic model in general 
gives a poorer fit to the TeV spectrum than the hadronic model. More 
detailed modeling is necessary to distinguish these emission models.

Using the MCMC method, we systematically investigate the parameter space
of models for the multi-wavelength emission of SNR RX J1713.7-3946. The
high quality of observational data, especially X-ray data from $Suzaku$
and TeV $\gamma$-ray data from $HESS$, enables us to get very good
constraints on most model parameters and better understand the emission
mechanisms. The radio and X-ray emissions are thought to be produced
by the synchrotron radiation of relativistic electrons accelerated in the
SNR. The high energy $\gamma$-ray emission (from GeV to TeV) can be produced
through the IC radiation of electrons scattering off background low
energy photons, and/or the decay of $\pi^0$ generated through CR
proton-ISM collisions. We study three kinds of scenarios, the leptonic
one, hadronic one and a hybrid one, distinguished through the emission
mechanism of high energy $\gamma$-rays. Thermal emissions, including
continuous and line emissions, are included in the modeling to constrain
density of the background plasma with the absence of thermal emission in
X-ray observations. 

The global fit of these three scenarios shows that:
1) the goodness-of-fit is the worst for the leptonic model and the best
for the hadronic model; 2) the X-ray data can set an upper limit on the
background ISM density, which is $0.03-0.009$ cm$^{-3}$ depending
on the temperature of the background electrons; 3) the upper limit of
$n_{\rm ISM}$ leads to a lower limit of energy content of relativistic
protons $W_p>4\times 10^{51}$ ($6\times 10^{51}$) erg for the hybrid
(hadronic) model, which seems to be too high for a typical core-collapse
supernova. The well constrained parameters of the leptonic scenario are
more physically acceptable, although the hybrid and hadronic scenarios
give smaller $\chi^2$ values. In some physically motivated models of
electron acceleration in SNR, the goodness-of-fit can be improved
\citep{2010A&A...517L...4F}, and our overall results appear to favor the
leptonic scenario \citep{2010ApJ...712..287E}. 
However, systematic errors in the X-ray and TeV band are 
significant in all the emission models studied so far, which demands more 
advanced modeling. For example, the observed source structure is much 
richer than the simple uniform zone assumed in all the emission models. 
On the other hand, given difficulties in X-ray and TeV observations, the 
errors of the relevant data may also be underestimated, especially the 
$HESS$ fluxes at the low energy range, which does not appear to smoothly 
match the recently obtained GeV fluxes with the {\it Fermi} observatory 
\citep{fermi:rxj1713}.

In the following we discuss several possible alternatives
of the hadronic scenario, in order to provide an independent judge of
the price needed to pay to keep the hadronic model working.
The most serious challenge to the hadronic and hybrid models is the
high energy content of relativistic protons inferred from the upper
limit of the background density. This high energy content is in
excess of typical supernova explosions, but could be explained by
hypernovae.  The total kinetic energy of the SNR with a shock speed
$U$ may be estimated as $$K={2\pi\over 3} R^3 n_{\rm ISM} m_p U^2
\simeq6\times 10^{50} \left(R\over 10\ {\rm pc}\right)^3\left(U\over
4500\ {\rm km}/{\rm s}\right)^2 \left(n_{\rm ISM}\over 0.03\ {\rm
cm}^{-3}\right) \rm ergs. $$ With the current best estimate of the
distance at 1\,kpc, the radius of the remnant is 10\,pc. For such a
remnant size, the total kinetic energy content ($K$) of the remnant is
much less than typical hypernovae and less than the energy content of
relativistic protons.  However, the distance to the SNR is not well
determined. The lack of thermal X-ray emission directly constrains the
integration along the line-of-sight of the thermal emissivity, which
is proportional to $n_{\rm ISM}^2 R$. The radius $R$ and shock speed
$U$ of the SNR scales linearly with the distance $D$. The upper limit
on $n_{\rm ISM}$ therefore is proportional to $D^{-1/2}$; $K$ is
proportional to $D^{9/2}$. The $\gamma$-ray flux produced via hadronic
process is proportional to $W_p n_{\rm ISM}/D^2$. The lower limit on
$W_p$ therefore is proportional to $D^{5/2}$. The total kinetic energy
of the SNR decreases more rapidly than $W_p$, making the discrepancy
between these two energies worse for shorter distance $D$. An increase 
of $D$ by a factor of 3, on the other hand,
can make the upper limit on $W_p$ less than $K$, which is now $\sim
8\times 10^{52}$ erg, at the high end of observed hypernovae
\citep{2005Ap&SS.298...81N}. This may explain the rareness of this
kind of shell type TeV SNRs with non-detection of thermal X-ray emission.
For the few tens of SNRs observed in X-rays, two of them show complete 
absence of thermal emission \citep{vink06}.
The efficiency of proton acceleration also needs to be greater than 
$50\%$ for this model to work \citep{2009Sci...325..719H}. The upper 
limit on the density is then about $0.02$ cm$^{-3}$, which is not too 
different from values obtained above and from previous studies.

Alternatively, the hadronic and hybrid models can work by overcoming
the upper limit on the ISM density placed by the upper limit on the
thermal electron emission. This limit arises from using the standard
ion-electron coupling term from \citet{2000ApJ...543L..61H} producing
a lower limit on the electron temperature (Eq. (1)). But this
assumed a standard profile for the remnant. If the emission of this
remnant is instead produced by the interaction of the supernova shock
striking a thin shell of material ejected from the star prior to
collapse (e.g. an outburst from a luminous blue variable or binary
mass ejection), the electrons might still be cold. However, to do so
would require very fine-tuned arguments (for the electron temperature
to be below $2\times10^6$K, the shock must have hit the stellar ejecta
less than 25\,yr ago). Even with this fine-tuning, this thin shell of
material would not have a large enough emitting volume to explain the
observed high-energy emission. A thin shell explanation seems unlikely.
It was also suggested that the effective particle acceleration would 
make the post-shock medium be less heated \citep{2009A&A...496....1D}, 
however, the extreme condition in such a scenario seems difficult to 
satisfy, and the non-linear effects will also make the medium be heated 
more significantly \citep{2010ApJ...712..287E}.

A plausible explanation of the low inferred density of the background 
plasma is the assumption of ionization equilibrium adopted in calculating 
the thermal emission, which however may not be valid.
For the case with a large fraction of neutral medium,
$n_e\ll n_{\rm ISM}$, the constraint on $n_{\rm ISM}$ from the thermal
emission, which is proportional to $n_en_{\rm ISM}$, will be much weaker
than the case with fully ionized gas. Therefore the requirement of $W_p$
can be smaller and the difference between $W_p$ and the power of typical
supernova can be reduced. In such a case, the low density of ionized
ISM implies that neutral gases can penetrate a depth of
$$ 6\times 10^{17} \left(n_{\rm ISM}\over0.03\ {\rm cm}^{-3}\right)
\left(U\over4500 {\rm km/s}\right) \left(T_e\over10^{7}{\rm K}
\right)^{0.23}{\rm cm}$$
before being ionized by the free electrons \citep{1978ApJ...225L..27C}.
For very low density of ionized plasma in the shocked downstream region,
this depth can be a significant fraction of the radius of the remnant,
and one may expect strong H$_\alpha$ emissions
\citep{2007ApJ...654L..69G, 2009Sci...325..719H}.
Better treatment of the ionization balance in the shock downstream may
address this issue.

Finally we mention that neutrino signal can be used to test the
hybrid/hadronic model of the $\gamma$-ray emission. It is shown
that if the TeV $\gamma$-rays are predominantly produced by hadronic
interactions, the accompanied neutrino signal might be detected
in the up-coming km$^3$ neutrino detector, such as KM3NET
\citep{2006PhRvD..74f3007K,2009A&A...495....9Y,2009APh....31..376M,
2010arXiv1010.1901Y}.

\acknowledgments

The MCMC code is adapted from the COSMOMC package \citep{cosmomc}. QY 
thanks Jie Liu for the help to develop the code. 
This work is supported in part by the Natural Sciences Foundation
of China (Nos. 10773011, 10963004, 11075169 and 11143007), the 973 project 
under the grant No. 2010CB833000, the Yunnan Provincial Science Foundation 
of China under the grant No. 2008CD061, a Marie Curie research fellowship 
at the University of Glasgow, and, under the auspices of the National 
Nuclear Security Administration of the U.S. Department of Energy at Los 
Alamos National Laboratory, supported by Contract No. DE-AC52-06NA25396.


\begin{thebibliography}{48}
\expandafter\ifx\csname natexlab\endcsname\relax\def\natexlab#1{#1}\fi

\bibitem[{Abdo} {et~al.}(2011)]{fermi:rxj1713}
{Abdo}, A. et al. 2011, arXiv:1103.5727

\bibitem[{Acero} {et~al.}(2009)]{2009A&A...505..157A}
{Acero}, F., {Ballet}, J., {Decourchelle}, A., {Lemoine-Goumard}, M., 
{Ortega}, M., {Giacani}, E., {Dubner}, G. \& {Cassam-Chena{\"i}}, G. 
2009, \aap, 505, 157

\bibitem[{Aharonian} {et~al.}(2004)]{2004Natur.432...75A}
{Aharonian}, F. et al. 2004, \nat, 432, 75

\bibitem[{Aharonian} {et~al.}(2005)]{aha05}
{Aharonian}, F. et al. 2005, \aap, 437, L7
  
\bibitem[{Aharonian} {et~al.}(2006)]{2006A&A...449..223A}
{Aharonian}, F., et al. 2006, \aap, 449, 223

\bibitem[{Aharonian} {et~al.}(2007)]{2007A&A...464..235A}
{Aharonian}, F., et al. 2007, \aap, 464, 235

\bibitem[{Aharonian} {et~al.}(2008)]{aha08}
{Aharonian}, F., et al. 2008, \apj, 692, 1500

\bibitem[{{Allen}(1973)}]{1973asqu.book.....A}
{Allen}, C.~W. 1973, {Astrophysical quantities} ({London: University of London,
  Athlone Press})

\bibitem[{{Axford}(1981)}]{1981ICRC...12..155A}
{Axford}, W.~I. 1981, in International Cosmic Ray Conference, Vol.~12, 155

\bibitem[{Benjamin} {et~al.}(2003)]{2003PASP..115..953B}
{Benjamin}, R.~A., et al. 2003, \pasp, 115, 953

\bibitem[{{Berezhko} \& {Ellison}(1999)}]{1999ApJ...526..385B}
{Berezhko}, E.~G. \& {Ellison}, D.~C. 1999, \apj, 526, 385

\bibitem[{{Berezhko} \& {V{\"o}lk}(2006)}]{2006A&A...451..981B}
{Berezhko}, E.~G. \& {V{\"o}lk}, H.~J. 2006, \aap, 451, 981

\bibitem[{{Berezhko} \& {V{\"o}lk}(2010)}]{2010A&A...511A..34B}
{Berezhko}, E.~G. \& {V{\"o}lk}, H.~J. 2010, \aap, 511, A34

\bibitem[{{Blasi}(2010)}]{2010MNRAS.402.2807B}
{Blasi}, P. 2010, \mnras, 402, 2807

\bibitem[{{Butt} {et~al.}(2008)}]{butt08}
Butt, Y. M., Porter, T. A., Katz, B., \& Waxman, E. 2008, \mnras, 386, L20

\bibitem[{Cassam-Chena{\"i}} {et~al.}(2004)]{2004A&A...427..199C}
{Cassam-Chena{\"i}}, G., {Decourchelle}, A., {Ballet}, J., {Sauvageot}, J. L.,
{Dubner}, G. \& {Giacani}, E. 2004, \aap, 427, 199

\bibitem[{{Chevalier} \& {Raymond}(1978)}]{1978ApJ...225L..27C}
{Chevalier}, R.~A. \& {Raymond}, J.~C. 1978, \apjl, 225, L27

\bibitem[{Drury} {et~al.}(2009)]{2009A&A...496....1D}
{Drury}, O'C., {Aharonian}, F.~A., {Malyshev}, D., \& {Gabici}, S. 2009, \aap,
  496, 1

\bibitem[{Ellison} {et~al.}(2010)]{2010ApJ...712..287E}
{Ellison}, D.~C., {Patnaude}, D.~J., {Slane}, P., \& {Raymond}, J. 2010, \apj,
  712, 287

\bibitem[{Enomoto} {et~al.}(2002)]{2002Natur.416..823E}
{Enomoto}, R., et al. 2002, \nat, 416, 823

\bibitem[{Fan} {et~al.}(2010{\natexlab{a}})]{2010MNRAS.406.1337F}
{Fan}, Z., {Liu}, S., \& {Fryer}, C.~L. 2010{\natexlab{a}}, \mnras, 406, 1337

\bibitem[{Fan} {et~al.}(2010{\natexlab{b}})]{2010A&A...517L...4F}
{Fan}, Z.~H., {Liu}, S.~M., {Yuan}, Q., \& {Fletcher}, L. 2010{\natexlab{b}},
  \aap, 517, L4

\bibitem[{Fang} {et~al.}(2009)]{2009MNRAS.392..925F}
{Fang}, J., {Zhang}, L., {Zhang}, J.~F., {Tang}, Y.~Y., \& {Yu}, H. 2009,
  \mnras, 392, 925

\bibitem[{Gamerman}(1997)]{Gamerman1997}
{Gamerman}, D. 1997, {Markov Chain Monte Carlo: Stochastic Simulation for
  Bayesian Inference} ({Chapman and Hall, London})

\bibitem[{Ghavamian} {et~al.}(2007)]{2007ApJ...654L..69G}
{Ghavamian}, P., {Laming}, J.~M., \& {Rakowski}, C.~E. 2007, \apjl, 654, L69

\bibitem[{Gabici}(2008)]{Gabici2008} {Gabici}, S. 2008, arXiv:0811.0836

\bibitem[{Helder} {et~al.}(2009)]{2009Sci...325..719H}
{Helder}, E.~A., et al. 2009, Science, 325, 719

\bibitem[{Hughes} {et~al.}(2000)]{2000ApJ...543L..61H}
{Hughes}, J.~P., {Rakowski}, C.~E., \& {Decourchelle}, A. 2000, \apjl, 543, L61

\bibitem[{Kamae} {et~al.}(2006)]{2006ApJ...647..692K}
{Kamae}, T., {Karlsson}, N., {Mizuno}, T., {Abe}, T., \& {Koi}, T. 2006, \apj,
  647, 692

\bibitem[{{Katz} \& {Waxman}(2008)}]{2008JCAP...01..018K}
{Katz}, B. \& {Waxman}, E. 2008, \jcap, 1, 18

\bibitem[{{Kistler} \& {Beacom}(2006)}]{2006PhRvD..74f3007K}
{Kistler}, M.~D. \& {Beacom}, J.~F. 2006, \prd, 74, 063007

\bibitem[{Koyama} {et~al.}(1995)]{Ko95}
{Koyama}, K., {Petre}, R., {Gotthelf}, E. V., {Hwang}, U., {Matsuura}, M.,
{Ozaki}, M. \& {Holt}, S. S. 1995, \nat, 378, 255

\bibitem[{Koyama} {et~al.}(1997)]{1997PASJ...49L...7K}
{Koyama}, K., {Kinugasa}, K., {Matsuzaki}, K., {Nishiuchi}, M., 
{Sugizaki}, M., {Torii}, K., {Yamauchi}, S. \& {Aschenbach}, B. 1997, 
\pasj, 49, L7

\bibitem[{Koyama} {et~al.}(2007)]{2007PASJ...59S..23K}
{Koyama}, K., et al. 2007, \pasj, 59, 23

\bibitem[{Lazendic} {et~al.}(2004)]{2004ApJ...602..271L}
{Lazendic}, J. S., {Slane}, P. O., {Gaensler}, B. M., {Reynolds}, S. P.,
{Plucinsky}, P. P. \& {Hughes} J. P. 2004, \apj, 602, 271

\bibitem[{Lewis} \& {Bridle}(2002)]{cosmomc}
{Lewis}, A. \& {Bridle}, S. 2002, \prd, 66, 103511

\bibitem[{Liu} {et~al.}(2008)]{2008ApJ...683L.163L}
{Liu}, S., {Fan}, Z., {Fryer}, C.~L., {Wang}, J., \& {Li}, H. 2008, \apjl, 683,
  L163

\bibitem[{{Mackay}(2003)}]{2003itil.book.....M}
{Mackay}, D.~J.~C. 2003, {Information Theory, Inference and Learning
  Algorithms} ({Cambridge University Press})

\bibitem[{Morlino} {et~al.}(2009{\natexlab{a}})]{2009MNRAS.392..240M}
{Morlino}, G., {Amato}, E., \& {Blasi}, P. 2009{\natexlab{a}}, \mnras, 392, 240

\bibitem[{Morlino} {et~al.}(2009{\natexlab{b}})]{2009APh....31..376M}
{Morlino}, G., {Blasi}, P., \& {Amato}, E. 2009{\natexlab{b}}, Astroparticle
  Physics, 31, 376

\bibitem[{{Moskalenko} {et~al.}(2006)}]{2006ApJ...640L.155M}
{Moskalenko}, I.~V., {Porter}, T.~A., \& {Strong}, A.~W. 2006, \apjl, 640, L155

\bibitem[{Muraishi} {et~al.}(2000)]{2000A&A...354L..57M}
{Muraishi}, H., et al. 2000, \aap, 354, L57

\bibitem[{Neal}(1993)]{Neal1993}
{Neal}, R.~M. 1993, {Probabilistic Inference Using Markov Chain Monte Carlo
  Methods} ({Department of Computer Science, University of Toronto})

\bibitem[{Nomoto} {et~al.}(2005)]{2005Ap&SS.298...81N}
{Nomoto}, K., {Maeda}, K., {Tominaga}, N., {Ohkubo}, T., {Deng}, J. 
\& {Mazzali}, P. 2005, Ap\&SS, 298, 81

\bibitem[{{Petrosian} \& {Liu}(2004)}]{PL04}
Petrosian, V. \& Liu, S. 2004, \apj, 610, 550

\bibitem[{{Plaga}(2008)}]{2008NewA...13...73P}
{Plaga}, R. 2008, New Astronomy, 13, 73

\bibitem[{Porter} {et~al.}(2006)]{2006ApJ...648L..29P}
{Porter}, T.~A., {Moskalenko}, I.~V., \& {Strong}, A.~W. 2006, \apjl, 648, L29

\bibitem[{{Raymond} \& {Smith}(1977)}]{1977ApJS...35..419R}
{Raymond}, J.~C. \& {Smith}, B.~W. 1977, \apjs, 35, 419

\bibitem[{Tanaka} {et~al.}(2008)]{2008ApJ...685..988T}
{Tanaka}, T., et al. 2008, \apj, 685, 988

\bibitem[{Uchiyama} {et~al.}(2003)]{2003A&A...400..567U}
{Uchiyama}, Y., {Aharonian}, F.~A., \& {Takahashi}, T. 2003, \aap, 400, 567

\bibitem[{Uchiyama} {et~al.}(2007)]{2007Natur.449..576U}
{Uchiyama}, Y., {Aharonian}, F.~A., {Tanaka}, T., {Takahashi}, T., \& {Maeda},
  Y. 2007, \nat, 449, 576
  
\bibitem[{Vink}(2006)]{vink06} 
{Vink}, J. 2006, arXiv:astro-ph/0601131

\bibitem[{{Vink} {et~al.}(2006)}]{2006ApJ...648L..33}
{Vink}, J., {Bleeker}, J., {van der Heyden}, K., {Bykov}, A., {Bamba}, A. 
\& {Yamazaki}, R. 2006, \apjl, 648, L33


\bibitem[{{Wang} {et~al.}(1997)}]{1997A&A...318L..59W}
{Wang}, Z.~R., {Qu}, Q., \& {Chen}, Y. 1997, \aap, 318, L59

\bibitem[{{Warren} {et~al.}(2005)}]{2005ApJ...634..376W}
{Warren}, J.~S., et al. 2005, \apj, 634, 376

\bibitem[{{Yamazaki} {et~al.}(2009)}]{2009A&A...495....9Y}
{Yamazaki}, R., {Kohri}, K., \& {Katagiri}, H. 2009, \aap, 495, 9

\bibitem[{{Yuan} {et~al.}(2010)}]{2010arXiv1010.1901Y}
{Yuan}, Q., {Yin}, P., \& {Bi}, X. 2010, ArXiv e-prints:1010.1901

\bibitem[{{Zirakashvili} \& {Aharonian}(2010)}]{2010ApJ...708..965Z}
{Zirakashvili}, V.~N. \& {Aharonian}, F.~A. 2010, \apj, 708, 965

\end{thebibliography}

\end{document}